\documentclass[natbib]{svjour3}

\usepackage[utf8]{inputenc}
\usepackage{amssymb,graphicx}

\renewcommand{\d}{\mathrm{d}}

\newcommand{\ii}{\mathrm{i}}

\title{Internal Cluster Structure}
\author{Matthias Bartelmann \and Marceau Limousin \and Massimo Meneghetti \and Robert Schmidt}
\institute
 {Universit\"at Heidelberg, Zentrum f\"ur Astronomie, Institut f\"ur Theoretische Astrophysik,
  Philosophenweg 12, 69120 Heidelberg, Germany \and
  Aix Marseille Universit\'e, CNRS, Laboratoire d'Astrophysique de Marseille, UMR 7326,
  13388, Marseille, France \and
  INAF -- Osservatorio Astronomico di Bologna, via Ranzani 1, 40127, Bologna, Italy and
  INFN -- Sezione di Bologna, viale Berti Pichat 6/2, 40127, Bologna, Italy \and
  Universit\"at Heidelberg, Zentrum f\"ur Astronomie, Astronomisches Rechen-Institut,
  M\"onchhofstra{\ss}e 12-14,
  69120 Heidelberg}

\date{draft version of \today}

\begin{document}

\maketitle

\begin{abstract}
The core structure of galaxy clusters is fundamentally important. Even though self-gravitating systems have no stable equilibrium state due to their negative heat capacity, numerical simulations find density profiles which are universal in the sense that they are fairly flat within a scale radius and gradually steepen farther outward, asymptotically approaching a logarithmic slope of $\approx-3$ near the virial radius. We argue that the reason for the formation of this profile is not satisfactorily understood. The ratio between the virial radius and the scale radius, the so-called concentration, is found in simulations to be closely related to the mass and the redshift and low for cluster-sized haloes, but observed to be substantially higher at least in a subset of observed clusters. Haloes formed from cold dark matter should furthermore be richly substructured. We review theoretical and observational aspects of cluster cores here, discuss modifications by baryonic physics and observables that can provide 
better 
insight into the internal structure of clusters.
\end{abstract}

\section{Fundamental importance of cluster cores}

%

\subsection{Difficulties with self-gravitating systems}

The core structure of galaxy clusters is of fundamental importance not only for astrophysics, but for physics as a whole. The essential reason for this statement is that the properties of the yet hypothetical dark-matter particles, and the dominance of not electromagnetically interacting dark matter itself, can be tested in the cores of galaxy clusters.

Due to the virial theorem, self-gravitating systems have a negative heat capacity. This is easily seen as follows. In its scalar version, the virial theorem demands that the total kinetic energy $E_\mathrm{kin}$ and the potential energy $E_\mathrm{pot}$ be related by
\begin{equation}
  2E_\mathrm{kin} = -E_\mathrm{pot}\;.
\label{eq:c-1}
\end{equation}
The factors multiplying the two forms of energy are due to the fact that the kinetic energy is a homogeneous function of degree $2$ of velocity, while the potential energy is a homogeneous function of degree $-1$ of position. The total energy is thus
\begin{equation}
  E = E_\mathrm{kin} + E_\mathrm{pot} = -E_\mathrm{kin}\;.
\label{eq:c-2}
\end{equation}
Self-gravitating systems can lose energy in various ways. For example, particles can become unbound and be ejected from the system in three-body encounters, or baryonic gas mixed with the dark matter can radiate energy and adiabatically contract the dark matter. If a self-gravitating system loses energy, no matter how, the system's total energy decreases. Since it is negative for a bound system, its absolute value increases thereby and, following the virial theorem (\ref{eq:c-2}), its kinetic energy must also increase. Cooling of the system by any kind of energy loss thus leads to heating of the system by contraction, which defines a negative heat capacity. This in turn implies that self-gravitating systems cannot reach a stable equilibrium state where a suitable thermodynamical potential could attain a minimum.

Nonetheless, numerical simulations of dissipation-less dark matter under the exclusive influence of gravity unanimously show that long-lived, gravitationally bound systems form in a wide range of masses, the so-called dark-matter haloes. Quite independent of various initial conditions such as the power spectrum of the dark matter fluctuations, these haloes turn out to have a universal density profile $\rho(r)$, characterised by two radii. Within the outer radius, the virial radius $R_\mathrm{vir}$ to be defined below, virial equilibrium is supposed to be achieved. Qualitatively speaking, the density profile is steep from an inner radius, the scale radius $r_\mathrm{s}$, to the virial radius, and flat from the halo centre to the scale radius. While simulations agree that the double-logarithmic slope
\begin{equation}
  \alpha = \frac{\d\ln\rho}{\d\ln r}
\label{eq:1}
\end{equation}
reaches the asymptotic value $-3$ near the virial radius, inner slopes between $-1$ and $-1.5$ are typically measured in simulated haloes. Introducing the dimension-less radius $x = r/r_\mathrm{s}$, density profiles of the functional form
\begin{equation}
  \rho(x) = \frac{\rho_\mathrm{s}}{x^\alpha(1+x)^{3-\alpha}}
\label{eq:2}
\end{equation} 
fit the numerically simulated dark-matter haloes very well, with $-1\lesssim\alpha\lesssim-1.5$. For $\alpha = -1$, the profile (\ref{eq:2}) was first suggested by \cite[][hereafter abbreviated by NFW]{NA97.1, NA96.1}, while \cite{MO98.3, MO99.2} favoured steeper central profiles with $\alpha \approx -1.4$. More recent, highly resolved numerical studies find that the density-profile slope decreases gently from the scale radius inward \citep{PO03.1, NA04.1} in a way that may depend on the halo mass \citep{JI00.1}. A density profile with an alternative functional form,
\begin{equation}
  \rho(x) = \rho_\mathrm{s}\exp\left[-2n\left(x^{1/n}-1\right)\right]\;,
\label{eq:2a}
\end{equation}
was suggested by Einasto \cite[see][]{EI89.1} and found to fit density profiles of numerically simulated dark-matter haloes very well \citep{ME06.1} with $n\approx5$. In (\ref{eq:2a}), the scale radius $r_\mathrm{s}$ was introduced such that the profile slope is $\alpha = -2$ at $r = r_\mathrm{s}$, which is also the case for the NFW profile at $x = 1$.

\begin{figure}[ht]
  \centering{\includegraphics[width=0.5\hsize]{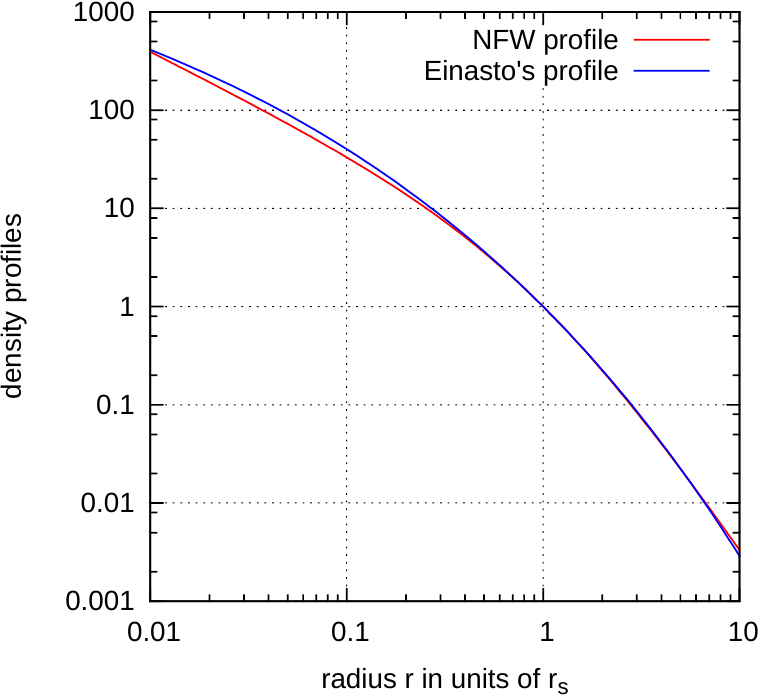}}
\caption{The NFW and Einasto density profiles (the latter with $n = 5$) are shown as functions of the dimension-less radial coordinate $x = r/r_\mathrm{s}$. The difference are subtle, but the Einasto profile's central slope gradually decreases to zero as the centre is approached. The profiles are scaled such that the density is unity at $r = r_\mathrm{s}$.}
\label{fig:1}
\end{figure}

The slope of the Einasto profile,
\begin{equation}
  \alpha = -2x^{1/n}\;,
\label{eq:2b}
\end{equation}
decreases continuously towards the centre and reaches zero there. One important difference between the generalised NFW profile (\ref{eq:2}) and the Einasto profile (\ref{eq:2a}), shown in Fig.~\ref{fig:1}, is the central density cusp in the former, but not in the latter. If the dark-matter particles decay or annihilate, the annihilation rates in both types of density profile may be substantially different. As two-body processes, annihilation rates are proportional to the squared particle density, hence cusped density profiles may predict higher rates than profiles with a finite-density core. More important than the central cusp, however, seems to be the level of substructure in cluster haloes, as we shall discuss later.

We shall return further below to the issue of the slope $\alpha$ near the centre of cluster haloes. For the purpose of our present discussion, the exact value of $\alpha$ is less important, however. Since stable final states of self-gravitating systems cannot be expected for the reasons given above, the mere fact that long-lived transient states like dark-matter haloes do exist at all and form a universal density profile is a most remarkable finding on its own.

The ratio between the two radii, the concentration parameter
\begin{equation}
  c = \frac{R_\mathrm{vir}}{r_\mathrm{s}}\;,
\label{eq:3}
\end{equation}
is found to be closely related to the masses and the formation times of the haloes. Generally, the earlier a halo forms, the larger the concentration parameter is. This is commonly attributed to the higher cosmic background density at the time of halo formation. We shall discuss the concentrations in more detail below.

\subsection{Hierarchical formation of dark-matter haloes}

\begin{figure}[ht]
  \centering{\includegraphics[width=0.5\hsize]{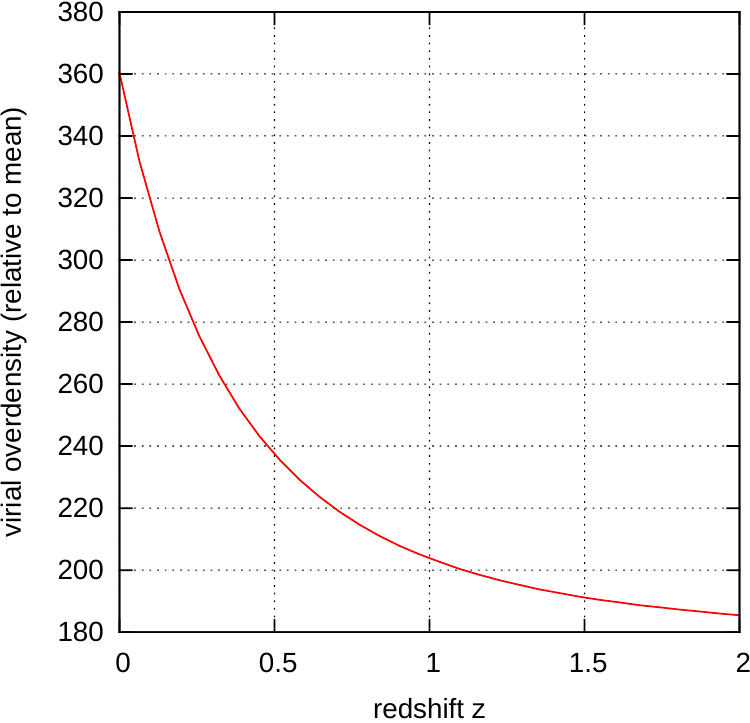}}
\caption{The virial overdensity $\Delta_\mathrm{v}$ relative to the mean cosmic density is shown here as a function of redshift for the standard cosmological model ($\Omega_\mathrm{m}=0.27$, $\Omega_\Lambda = 1-\Omega_\mathrm{m}$).}
\label{fig:5}
\end{figure}

The global properties of a dark-matter halo, i.e.~its radius $R_\mathrm{vir}$ and its mass $M$, are commonly related to each other and to the parameters of the cosmic background through the spherical collapse model. Specifically, the model states that the mean matter density enclosed by the virial radius should be higher than the mean cosmic density $\bar\rho$ by the virial overdensity $\Delta_\mathrm{vir}$, thus
\begin{equation}
  \frac{3M}{4\pi R_\mathrm{vir}^3} = \Delta_\mathrm{vir}\bar\rho\;.
\label{eq:4}
\end{equation}
The virial overdensity $\Delta_\mathrm{vir}$ depends on redshift and on cosmology. In terms of the matter-density parameter $\Omega_\mathrm{m0}$ at the present epoch or of the matter-density $\Omega_\mathrm{m}(z)$ as a function of redshift, the mean cosmic matter density is
\begin{equation}
  \bar\rho(z) = \frac{3H_0^2}{8\pi G}\Omega_\mathrm{m0}(1+z)^3 =
  \frac{3H^2(z)}{8\pi G}\Omega_\mathrm{m}(z)\;.
\label{eq:5}
\end{equation} 
For an Einstein-de Sitter universe with critical matter density and without a cosmological constant, the mean overdensity is analytically found to be $\Delta_\mathrm{vir} = 18\pi^2 \approx 178$. For the standard cosmological model, $\Delta_\mathrm{v}$ (relative to the mean cosmic density) is substantially higher and changes with redshift, as shown in Fig.~\ref{fig:5}. Often, $\Delta_\mathrm{v}$ is instead given relative to the critical cosmic density and thus reduced by a factor $\Omega_\mathrm{m}(z)$ compared to the values shown in Fig.~\ref{fig:5}.

The considerations so far show that the formation and in particular the existence of long-lived transient states of dark-matter haloes, at least in numerical simulations of collision-less dark matter, seems to be a remarkably well-determined process without much freedom. The collapse dynamics sets the virial overdensity $\Delta_\mathrm{vir}$ which, together with the cosmic mean matter density $\bar\rho$, relates the virial radius to the mass. The concentration, which itself seems to be defined by the formation redshift, then sets the scale radius. Given that, the density profile is defined.

If the dark matter is cold, the power spectrum $P(k)$ of the initial density fluctuations from which the haloes ultimately form has a characteristic shape that is again fully determined by few cosmological conditions. For small $k$, it increases as a power law in $k$, $P(k)\propto k^n$ with $n\lesssim1$. For very large $k$, it decreases asymptotically like $P(k)\propto k^{n-4}$. The power-law index $n$ is set by cosmological inflation very early during the evolution of the universe. If inflation had lasted eternally, the power spectrum would be precisely scale-free, $n = 1$. Since inflation must have ended after a finite time, $n$ should be slightly smaller. It belongs to the most exciting results of the WMAP satellite mission that $n$ is indeed slightly, but significantly smaller than unity, $n = 0.968\pm0.012$ \citep{KO11.1}.

The decrease of the slope by $-4$ is also entirely defined by the inhibition of growth of dark-matter fluctuations inside the horizon during the radiation-dominated epoch. The maximum of the power spectrum is thus set by the horizon size at the epoch of matter-radiation equality, which is in turn defined by the cosmic matter-density parameter. It is essential for our purposes to note that the dark-matter fluctuation power spectrum has no free parameters except for its overall normalisation. Traditionally, this normalisation is fixed by the parameter $\sigma_8$, generally defined by specialising the scale-dependent variance
\begin{equation}
  \sigma_R^2 = \int_0^\infty\frac{k^2\d k}{2\pi^2}\,P(k)\,|\hat W_R(k)|^2
\label{eq:6}
\end{equation}
to the radius $R = 8\,h^{-1}\,\mathrm{Mpc}$. In (\ref{eq:6}), $\hat W_R(k)$ is a Fourier-transformed window function introduced to smooth fluctuations on scales smaller than $R$. The specific setting of this length scale to $8\,h^{-1}\,\mathrm{Mpc}$ has historical reasons: First measurements of $\sigma_8$ based on the galaxy distribution found that variance of the density contrast was approximately unity on a scale of $8\,h^{-1}\,\mathrm{Mpc}$ \cite[cf.][for an example]{DA83.1}.

The power in the dark-matter fluctuations is proportional to the volume in $k$-space times the power spectrum, or $\propto k^3P(k)$, which asymptotically tends to $k^{n-1}$ for large $k$. Since $n\lesssim1$, the power keeps rising with $k$ until it gently flattens off asymptotically for very large $k$. This means that the largest power is in the smallest fluctuations, giving rise to the bottom-up scenario of cosmological structure formation in cold dark matter: Those fluctuations with the largest power form first, hence structures in cold dark matter start forming at the smallest scales. It is thus expected that cold dark matter clumps on all scales. Moreover, the largest structures will form the latest. Depending on the normalisation parameter $\sigma_8$, a maximum mass is expected which could just have formed in the cosmic present.

The statistics of cosmological structures is characterised by the assumption that they form from an initially Gaussian random field with a low amplitude. During the linear phase of structure growth, this amplitude increased proportional to the linear growth factor $D_+(a)$ without any interaction between different Fourier modes of the density field. The spherical-collapse model predicts that non-linear collapse concludes when the linear density contrast has reached a critical value $\delta_\mathrm{c}$ which is
\begin{equation}
  \delta_\mathrm{c} = \frac{3}{5}\left(\frac{3\pi}{2}\right)^{2/3} \approx 1.686
\label{eq:7}
\end{equation} 
in an Einstein-de Sitter universe and not much different in other model universes. The abundance of cosmological objects of a certain mass can then be derived directly from the peak statistics in a Gaussian random field. The result is the Press-Schechter mass function,
\begin{equation}
  N(M, a) = \sqrt{\frac{2}{\pi}}\frac{\bar\rho_0\delta_\mathrm{c}}{\sigma_RD_+(a)}
  \frac{\d\ln\sigma_R}{\d M}
  \exp\left(-\frac{\delta_\mathrm{c}^2}{2\sigma_R^2D_+^2(a)}\right)\frac{\d M}{M}\;,
\label{eq:8}
\end{equation}
where $\bar\rho_0$ is the mean cosmic matter density today. The radius $R$ and the mass $M$ are related by requiring that a sphere with radius $R$ filled with the mean cosmic matter density should contain the mass $M$,
\begin{equation}
  \frac{4\pi}{3}R^3\bar\rho = M\;.
\label{eq:9}
\end{equation}
Again, the details of this mass function and its possible modifications are not relevant for our discussion. It is important, however, that the mass function cuts off exponentially at a mass scale defined by
\begin{equation}
  D_+(a)\,\sigma_R \approx \delta_\mathrm{c}\;.
\label{eq:10}
\end{equation} 
The larger $R$ is, and therefore the larger the masses are that are being considered, the lower is the variance $\sigma_R$ and the larger $D_+(a)$ has to become for (\ref{eq:10}) to be satisfied. This quantifies the earlier statement that the most massive objects form late in cosmic history.

\begin{figure}[ht]
  \includegraphics[width=0.49\hsize]{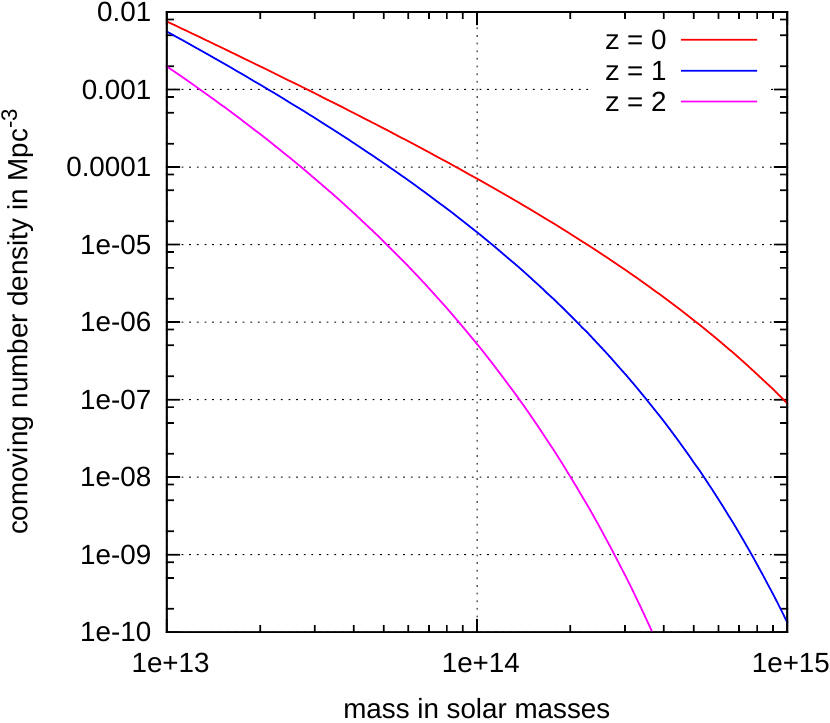}\hfill
  \includegraphics[width=0.49\hsize]{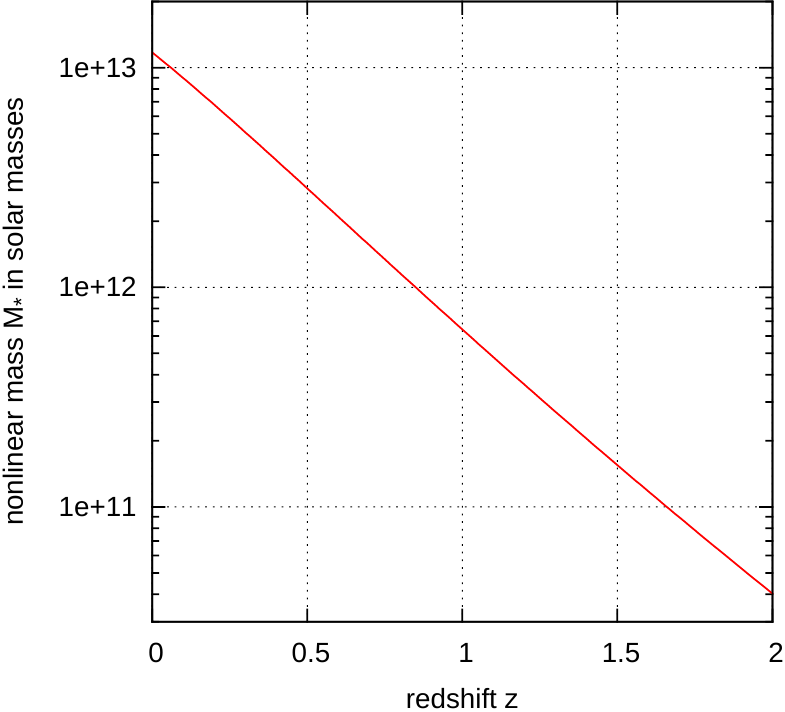}
\caption{\textit{Left panel}: The Press-Schechter mass function is shown as a function of mass for three different redshifts. The number density of massive clusters is evolving very rapidly as a function of time, showing that clusters are a dynamical population of cosmic objects. \textit{Right panel}: The non-linear mass $M_*$ is shown as a function of redshift for a standard $\Lambda$CDM cosmological model. Even at redshift zero, the non-linear mass is just above $10^{13}\,M_\odot$, showing that clusters with their masses $\gtrsim10^{14}\,M_\odot$ are highly non-linear objects.}
\label{fig:2}
\end{figure}

A typical non-linear mass scale is commonly defined by combining Eqs.~(\ref{eq:10}) and (\ref{eq:6}). Define a window function $W_R(r)$, for example a top-hat function $W_R(r) = \Theta(R-r)$. Its Fourier transform
\begin{equation}
  \hat W_R(k) = \frac{1}{2\pi^2}\frac{\sin kR}{kR}\;,
\label{eq:6a}
\end{equation} 
inserted into (\ref{eq:6}), returns the variance $\sigma_R^2$ on the scale $R$. As $R$ increases, progressively more structure is smoothed out, so $\sigma_R$ decreases. Find the scale where $\sigma_R = \delta_\mathrm{c}$ and convert this to a mass through Eq.~(\ref{eq:9}), which is called the non-linear mass $M_*$. The non-linear mass for a standard $\Lambda$CDM cosmological model is shown in the right panel of Fig.~\ref{fig:2}.

\subsection{Expected level of substructuring}

Numerical simulations at high resolution, assuming cold dark matter, find strongly substructured haloes. Numerous different simulation runs undertaken during recent years arrive at the following statements on the substructure level:

The amount of substructure in a halo increases with the mass of the host halo. At fixed host-halo mass, it decreases with the concentration of the host halo and with its formation redshift. These trends are noticeable, but of comparable amplitude as the scatter in the subhalo abundance between similar host haloes. The abundance of low-mass substructures per unit host-halo mass reflects that of the surrounding Universe as a whole. The subhaloes are substantially less spatially concentrated than the dark matter. Different approaches to the identification of subhaloes within host haloes lead to similar substructure abundance functions \citep{GA04.2, ZE05.1, GI10.2}.

Taking into account baryons in addition to the dark matter, \cite{DO09.2} find that the diffuse hot gas is efficiently removed from subhaloes when they enter their host haloes. Compared to simulations with purely dark matter, stripping of the gas may decrease the subhalo mass function. However, if radiative cooling of the baryons is taken into account, baryons condense towards the centre of the subhaloes and form compact stellar cores, thus mitigating the effect of stripping. In all variants of the simulations including baryons, only about $1\,\%$ of the subhaloes within the virial radii of their host haloes preserve a gravitationally bound hot gaseous atmosphere.

Based on the Millennium-II simulation data, \cite{BO09.1} proposed a fitting formula for the abundance function of subhaloes,
\begin{equation}
  N(\mu) = \left(\frac{\mu}{0.01}\right)^{-0.94}\exp\left[-\left(\frac{\mu}{0.1}\right)^{1.2}\right]\;,
\label{eq:6b}
\end{equation} 
where $\mu = (M_\mathrm{sub}/M_{200})$ is the ratio between the masses of subhaloes and the virial mass of their host halo. \cite{GA11.1} confirm that the fractional mass in subhaloes, $\mu N(>\mu)$, gently increases until $\mu\approx0.01$ and falls steeply above. The peak in the fractional mass increases from $\approx0.1\,\%$ to $\approx1\,\%$ from galaxy- to cluster-sized host haloes.

\subsection{Dark-matter annihilation?}

Recent observations and surveys of high-energy gamma rays with imaging Cherenkov telescopes and with the Large Area Telescope (LAT) on-board the Fermi satellite have resulted in a variety of theoretical investigations of galaxy clusters and interesting inferences. So far, galaxy clusters have escaped detection both by Cherenkov imaging and by Fermi-LAT. Both types of telescope cover very wide ranges of gamma energies. While the High-Energy Stereoscopic System (H.E.S.S.) is sensitive to gamma photons with energies between $\sim10\,\mathrm{GeV}$ to $\sim10\,\mathrm{TeV}$, Fermi-LAT shows the highest response at smaller gamma energies between $\sim1\,\mathrm{GeV}$ and $\sim100\,\mathrm{GeV}$.

After one year of operation, the non-detection with Fermi-LAT of a gamma signal from nearby galaxy clusters led \cite{YU10.1} to conclude that cluster substructures should have masses of at least $M\ge 10^2\ldots10^3\,M_\odot$. The argument behind this conclusion is straightforward: Substructures in clusters are expected to be substantially denser than the surrounding, smoothly distributed cluster matter, and are thus expected to emit a good fraction, if not most of a cluster's hypothetical gamma signal. The non-detections of gamma rays from nearby clusters then imply an upper limit to the lumpiness of cluster matter. Assuming a mass function for the cluster substructures, this upper limit translates to a lower limit of substructure masses, since the mass function should increase quite steeply towards decreasing mass. \cite{SA11.1} used essentially the same line of reasoning to argue that nearby galaxy clusters may be more efficient probes of dark-matter annihilation than local dwarf galaxies when viewed 
with 
imaging Cherenkov telescopes. Conversely, \cite{PI11.1} argued that, should clusters contain substructures down to the Earth's mass, models of dark-matter annihilation through leptonic channels (so-called leptophilic models) would be ruled out by the lack of detections.

Similarly, the cross section for dark-matter annihilation was constrained from above by the gamma non-detections. \cite{AN12.1} used 2.8 years of Fermi-LAT data taken from 49 nearby clusters to derive $\langle\sigma v\rangle\lesssim10^{-22}\mathrm{cm^3s^{-1}}$ at $10^3\,\mathrm{GeV}$ for the velocity-averaged annihilation cross section without taking into account that baryonic energy dissipation may contract the dark-matter distribution, and argued that baryonic contraction could lower this upper limit by about an order of magnitude. From three years of Fermi-LAT data and gamma non-detection from eight nearby galaxy clusters, \cite{HU12.1} derived a lower limit on the dark-matter decay time of $\tau\gtrsim10^{26}\,\mathrm{s}$ and an upper limit on the dark-matter annihilation cross section of $\langle\sigma v\rangle\lesssim10^{-22}\mathrm{cm^3s^{-1}}$ at gamma energies near $10^3\,\mathrm{GeV}$. These constraints tighten by $\sim3$ orders of magnitude once cluster substructures are taken into account. \cite{
AB12.1} placed an upper limit on the dark-matter annihilation cross section of $\langle\sigma v\rangle\lesssim10^{-23}\mathrm{cm^3s^{-1}}$ at $10^3\,\mathrm{GeV}$, possibly lowering by $\gtrsim2$ orders of magnitude due to substructure, from the non-detection of the nearby Fornax cluster in 14.5 hours of observation with H.E.S.S.

These conclusions are well consistent with each other and show that upper limits on the annihilation cross section as tight as $\langle\sigma v\rangle\lesssim10^{-25}\mathrm{cm^3s^{-1}}$ can be inferred at gamma energies near $\sim1\,\mathrm{TeV}$ from the gamma non-detections if galaxy clusters are substructured as expected from simulations. Interestingly, \cite{GA12.1} deduce from highly resolved simulations not only that the dark-matter annihilation signal from galaxy clusters should be stronger than from galaxies because of the substructures, but also that it should emerge preferentially from large cluster-centric radii since substructures are likely to be tidally destroyed near cluster cores. They identify the Coma and Fornax clusters as most promising candidates for gamma detections in future observations.

These results should be read with some caution, however. Firm predictions of the gamma emission expected from galaxy clusters require that the abundance and the central density of substructures with very low masses be accurately known, possibly down to stellar- or even lower-mass structures. This is well outside the reach of even the best-resolved numerical simulations today. Current inferences from the non-detection of gamma rays from clusters are thus based on substantial extrapolations of simulation results towards masses well below their resolution limit.

Summarising the discussion so far, we see that galaxy clusters, which are at the massive end of the spectrum of gravitationally bound cosmic structures, formed latest during the cosmic evolution if structure formation proceeded in a bottom-up way, as expected if the dark matter is cold. In such a scenario, the most massive objects must always be a dynamically active population since they are the most recent to have formed and did at least in part not have the time to relax. Small, lower-mass structures formed earlier, which means that the more massive objects must at least partially have been assembled by merging smaller units. Even though it is expected that part of the small structures that finally end up as parts of a galaxy cluster get disrupted, clusters should host sublumps with a broad continuum of masses. The mass fraction contained in the sublumps is expected to reach the per cent level, with subhaloes of approximately one per cent of the cluster mass contributing most. Tidal disruption of subhaloes 
near cluster cores renders the spatial subhalo distribution substantially flatter than that of the dark matter. The non-detection of gamma rays from clusters puts firm upper limits on the dark-matter annihilation cross section and on the decay time of the dark-matter particles.

\begin{figure}[ht]
  \centering{\includegraphics[width=0.5\hsize]{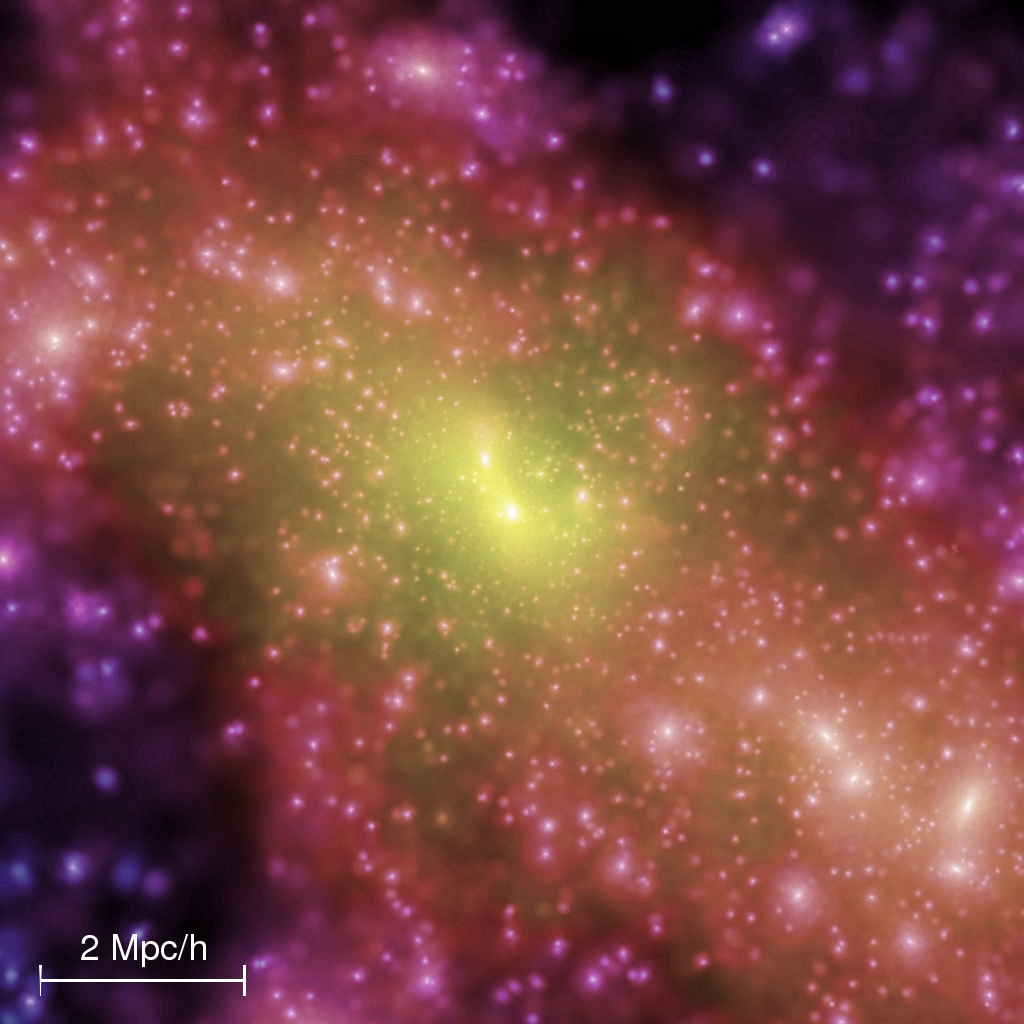}}
\caption{Haloes made of cold dark matter are expected to contain rich substructure on all scales. (Image extracted from the Millennium simulation, \citealt{SP05.1})}
\label{fig:3}
\end{figure}

\subsection{Attempts at understanding universality}

\begin{figure}[ht]
  \centering{\includegraphics[width=0.5\hsize]{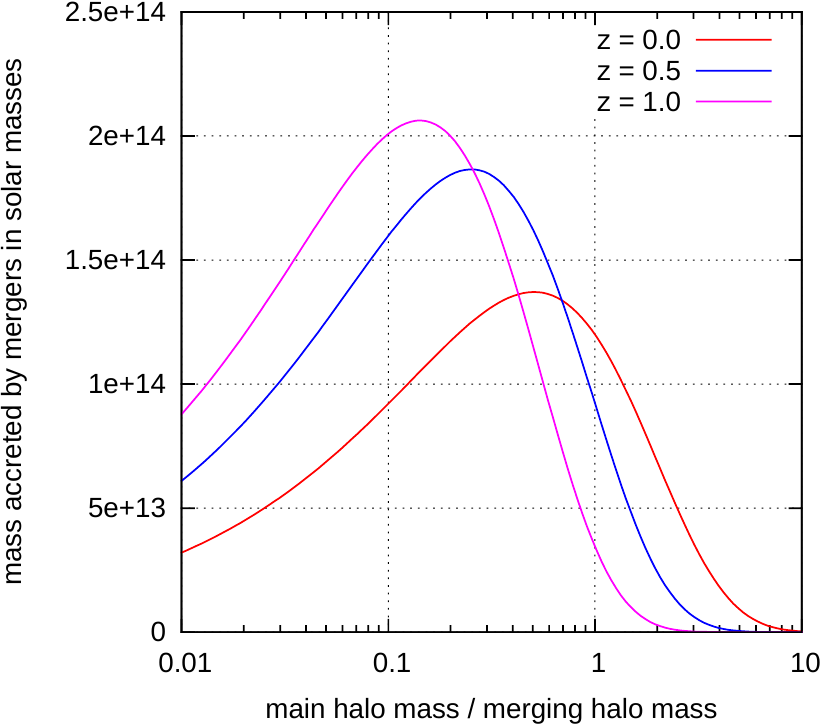}}
\caption{Amount of mass accreted by a halo of mass $5\times10^{14}\,M_\odot$ per unit logarithmic scale-factor as a function of the ratio between the masses of the main halo and the merging haloes. At redshift zero, most of the mass is accreted in lumps having about half of the main-halo mass. At higher redshift, most of the mass comes in smaller pieces.}
\label{fig:4}
\end{figure}

As we have described in some detail above, clusters as well as their substructures are expected to have universal density profiles according to numerical simulations. Despite many attempts to understand which physical processes shape this profile in absence of an equilibrium state, it is probably fair to say that a fundamental reason has not been found yet. Two main theoretical concepts have so far been discussed as possible explanations. One relies on the assembly of haloes from smaller haloes and the importance of the merging processes and rates involved \citep{SY98.1, SU00.1}. If $n$ is the slope of the dark-matter power spectrum near the scale corresponding to the halo mass considered, then halo density profiles should be power laws with slope $-3(3+n)/(5+n)$. For $n\approx-2$ as appropriate for cluster-sized haloes, slopes near $-1$ can be achieved, as simulations find them near the cores of dark-matter haloes. Since the universal profile is found for haloes with masses spanning several orders of 
magnitude, haloes of galaxy mass for which $n$ approaches $-3$ should have flatter cores than haloes of cluster mass, which is contrary to other claims based on simulations \citep{JI00.1}. Moreover, halo mergers are important for this scenario, which are frequent in any bottom-up hierarchical scenario of structure formation such as cold dark matter (see Fig.~\ref{fig:4}). Simulations undertaken with hot dark matter, in which structure formation proceeds from the top down and mergers are rare, however, produce haloes with the same universal density profile, but with an inverted concentration-mass relation. Mergers can therefore not be essential for the formation of the universal profile \citep{MO99.2, WA09.1}.

Another extensively discussed concept relies on the secondary-infall model \cite[see][for a recent example]{AS07.1}, which essentially rests on the fact that the so-called radial action is an adiabatic invariant. Generally, for a mechanical system with a Hamiltonian function varying slowly compared to the orbital time, the quantity
\begin{equation}
  I = \oint\,\vec p\cdot\d\vec q
\label{eq:10a}
\end{equation}
is invariant in the adiabatic approximation, where $\vec q$ and $\vec p$ are the vectors of the generalised coordinates and their conjugate momenta. Particles of mass $m$ on circular orbits in the potential of a spherically-symmetric mass $M(r)$, for example, have circular velocities
\begin{equation}
  v_\mathrm{c}(r) = \sqrt{\frac{GM(r)}{r}}\;.
\label{eq:10b}
\end{equation} 
For their orbits, therefore, the quantity
\begin{equation}
  I = 2\pi m\sqrt{GM(r)r}
\label{eq:10c}
\end{equation} 
is invariant while the potential changes adiabatically. For matter orbiting in existing potential wells that deepen slowly by accreting matter from their environment, the square root of the radius times the mass is adiabatically invariant. While this assumption yields density profiles approximating the universal profile, the scatter due to varying initial conditions and angular momenta is large. Besides, orbital times are of order
\begin{equation}
  \tau_\mathrm{orbit} \approx \frac{r}{v_\mathrm{c}(r)} \approx
  \frac{r^{3/2}}{\sqrt{GM(r)}}\;,
\label{eq:11}
\end{equation}
while the time scale for changes in the potential may be as fast as the free-fall time scale,
\begin{equation}
  \tau_\mathrm{ff} \approx \frac{1}{\sqrt{G\rho}}\;,
\label{eq:12}
\end{equation}
which is of the same order as the orbital time. This sheds doubt on the adiabatic assumption at least in such circumstances when rapid mergers change the potential on a short time scale comparable to the free-fall time scale. Other discussions begin by postulating a certain phase-space structure of the dark-matter halo particles, for example a power law in the ratio $\rho/\sigma^3$ between the density and the velocity dispersion $\sigma$ \citep{AU05.1}. Among others, the Jeans equation then admits constrained solutions which may come near the universal profile, but then the origin of the initial phase-space structure remains unexplained.

Of course, this brief discussion cannot do justice to the many efforts to explain the fundamental problem why a physical process without an equilibrium state should reveal universal behaviour, but it may serve to illustrate why the mere existence of a universal profile, quite irrespective of its details, raises conceptually important questions that have so far escaped a fundamental explanation.

\subsection{Halo concentrations and triaxiality}

Comparably fundamental may be the concentration parameter which controls the shape of the universal profile. Simulations show that it should decrease with mass quite slowly, $c\propto M^{-0.1}$ or similar, and more steeply on redshift. It should be mentioned that there is at least one simulation \citep{PR12.1} claiming that the concentration-mass relation be non-monotonic, increasing towards high masses. At the time of writing of this review, it is yet unclear why this result disagrees with virtually all other simulations of dark-matter haloes, and whether it may be seen as an artefact of interpretation.

\begin{figure}[ht]
  \includegraphics[width=0.49\hsize]{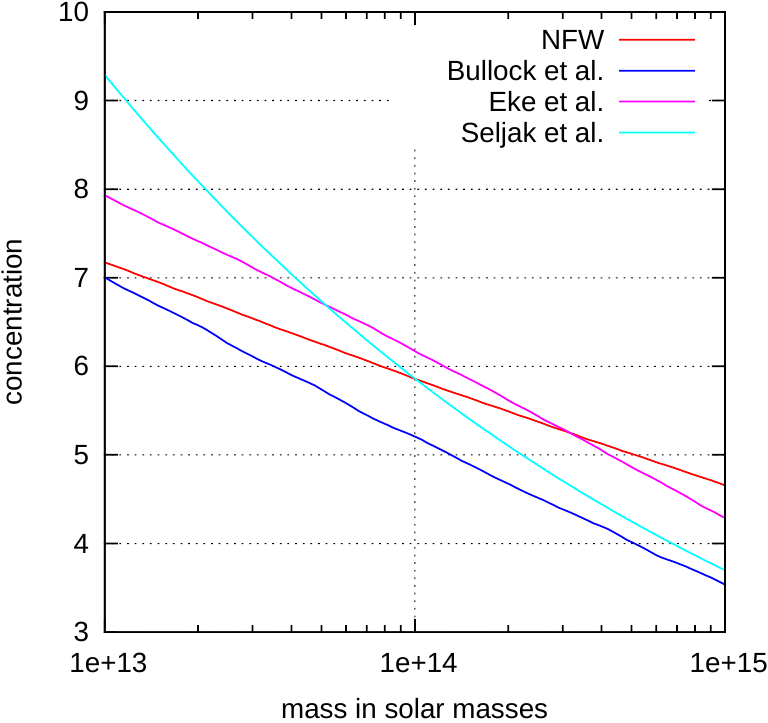}\hfill
  \includegraphics[width=0.49\hsize]{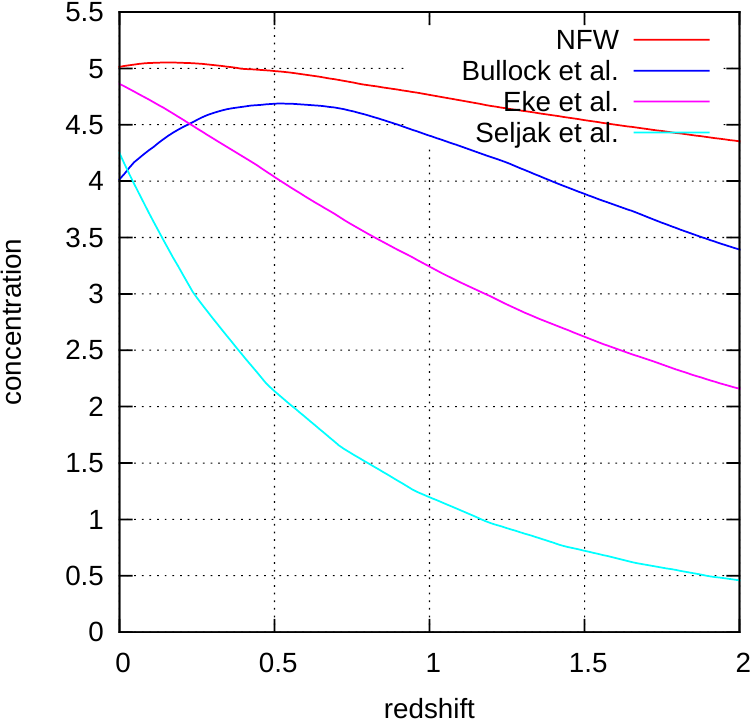}
\caption{\textit{Left panel}: Examples of concentration-mass relations at redshift zero in the mass range of galaxy cluster according to different algorithms relating the formation time to the halo core density. \textit{Right panel}: Evolution of the concentration of haloes with mass $5.0\times10^{14}\,M_\odot$ with redshift according to the same algorithms. (\citealt{BU01.1, EK01.1, NA97.1, NA96.1, SE00.1}; see \citealt{DO04.2, DU08.1, GA08.1, MA08.2, MA07.5, NE07.1, SH06.2, ZH09.1} for some other examples)}
\label{fig:6}
\end{figure}

The decrease of the concentration with mass is generally explained as a reflection of halo formation time. As mentioned above, it is assumed that halo cores preserve the mean cosmic density at the time of their formation, which at least qualitatively reproduces the trend seen in simulations of hierarchical models of structure formation such as cold dark matter: Massive haloes form later and should thus have lower concentrations. This is supported by the finding that the concentration-mass relation is inverted in top-down formation scenarios occurring in warm dark-matter models \citep{WA09.1}.

As is detailed elsewhere (see the review by Ettori et al. in this volume), some measured concentrations disagree strongly with those expected from the concentrations-mass relations found in numerical simulations \citep{BR05.1, BR08.3, CO07.1, HE07.1, LE09.1, UM08.1, UM10.1}. In particular, concentrations derived from strong gravitational lensing tend to be substantially and significantly higher for massive clusters, sometimes more than twice as high. For massive clusters, concentrations around $c = 5$ should be found, as Fig.~\ref{fig:6} shows. At fixed mass, concentrations of simulated haloes have a log-normal distribution with a standard deviation of $\approx 0.2$, corresponding to typical relative concentration fluctuations of $\approx 20$ per cent around the mean. Concentrations as high as those routinely found in strongly-lensing clusters are difficult to understand since they can hardly be accommodated with the standard $\Lambda$CDM cosmological model. Baryonic physics and feedback processes in cluster 
cores may affect the expected concentrations, though, as discussed by \cite{ME10.3}. They find that higher concentrations occur if baryonic cooling is taken into account, while AGN feedback brings the concentrations back near the range expected without baryons.

However, a substantial concentration bias can be expected in galaxy-cluster samples selected for their strong-lensing effects. Selecting strongly gravitationally lensing cluster-sized haloes in simulations, \cite{ME10.1} find that such clusters have concentrations tend to have concentrations higher by $\approx50\,\%$ than average. Taking this bias into account, \cite{UM11.1} find that their high concentrations derived from strong and weak lensing in four similar clusters maybe compatible with theoretical expectations.

However, high concentrations in clusters are not an unanimous result. Some strong-lensing studies, 
sometimes even of the same clusters from which high concentrations had been inferred, find concentrations in the expected range \citep{HA06.1, LI08.1}. In a strong-lensing analysis of 12 clusters at redshift $z > 0.5$ from the Massive Clusters Survey (MACS), \cite{SE12.1} find good agreement of the measured concentrations with theoretical expectations if the possible triaxiality of the cluster mass distribution is taken into account, even though their effective Einstein radii are still substantially higher than expected for $\Lambda$CDM. To further complicate the picture, concentrations inferred from weak gravitational lensing generally seem to reproduce the expected concentrations well. To give just two recent and possibly extreme examples, \cite{LE11.1} find a concentration of $c = 4.0^{+14}_{-2}$ from a weak-lensing analysis of the cluster XMMU J1230.3$+$1339 at $z = 0.975$. \cite{OK10.1} analyse the concentration-mass relation with weak gravitational lensing in 30 X-ray luminous clusters. They find a 
fairly steep dependence of the concentrations on mass, $c \propto M^{-0.40\pm0.19}$, but the inferred concentration of $c = 3.48^{+1.65}_{-1.15}$ for clusters with a virial mass near $M = 10^{15}\,h^{-1}\,M_\odot$ are just as expected. Yet another result is found by \cite{MA08.1}, who find concentrations lower than expected in a large sample of haloes from a wide mass range analysed by weak lensing. Using numerical simulations, \cite{BA12.1} find that the concentrations expected to be measured from weak lensing should be slightly biased low.

In contrast to gravitational lensing, X-ray observations \citep[e.g.][]{BU07.1, ET11.1, ET10.1, GA07.3, PO05.1, SC07.4} and interpretations of galaxy kinematics in clusters \cite[e.g.][]{RI06.1, WO10.1} typically find good agreement between measured and expected concentrations, in particular for clusters considered relaxed, albeit with considerable scatter. Detailed analyses of cluster cores based on X-ray data or on lensing data combined with the stellar dynamics in the central cluster galaxy find a large spread of density profiles, ranging from noticeably steeper than NFW \citep{LE03.1, BU04.1} to considerably flatter \citep{SA04.1, NE13.1, ET02.1}. Clearly, no final word can yet be said on the central density profiles of clusters. An outstanding example of a cluster showing an NFW density profile is Abell~2261 \citep{CO12.1}, observed within the CLASH programme \citep{PO12.1}.

Many recipes exist giving empirical descriptions of the concentration-mass relation. A selection of them is displayed in Fig.~\ref{fig:6}. Typically, they define a halo-formation redshift as the redshift when a certain fraction of the final halo mass was assembled, and relate the halo core density to the mean cosmic density at that redshift. Several of these recipes \citep{NA97.1, BU01.1, EK01.1, NE07.1, SE00.1, ZH09.1} are found to fit the mass-concentration relations of simulated haloes quite well, at least in certain wide mass and redshift ranges. However, the formation redshifts defined that way have to be very high for the simulated concentration-mass relation to be explained. Apparently, not more than a few per cent of the final halo mass need to be assembled when the final core density of the forming halo is imprinted. At the very least, this is an astonishing finding that should ultimately be understood.

Another difficulty related to the halo concentrations has to do with the so-called halo model of non-linear cosmic structure formation. It assumes that all cosmic structures are composed of haloes whose two-point correlation defines the power spectrum on large scales, while the small-scale power spectrum is given by the mass auto-correlation inside individual haloes. This being so, the internal density structure of individual haloes is an important ingredient of the halo model. However, for reproducing the non-linear, small-scale power spectrum found in numerical simulations, concentration-mass relations typically have to be assumed which are substantially steeper functions of mass than found by direct measurement in the very same simulation data. It seems that this substantial discrepancy can be solved by taking into account that haloes are substructured and thus contain a hierarchy of lower-mass haloes \citep{GI10.1}.

It can be shown analytically \citep{DO70.1} that cosmic structure formation by gravitational collapse from an initially Gaussian random field has to proceed in an anisotropic way. The probability for any pair of eigenvalues of the so-called Zel'dovich deformation tensor to be equal vanishes identically, hence structures have to evolve towards triaxial bodies at least well into the mildly non-linear regime. It is thus natural to expect that the shape of dark-matter haloes should approximate triaxial ellipsoids characterised by two axis ratios, called the ellipticity and the prolaticity, for the statistical distribution of which any structure-formation scenario based on an Gaussian initial conditions makes specific predictions. Modelling of strong gravitational lensing (see there) shows that triaxial haloes can better fit the gravitational-lensing effects simultaneously with the X-ray emission of galaxy clusters, and the axis ratios found that way are generally in a reasonable range.

The brief review given so far shows that the core structure of galaxy clusters poses a number of substantial problems, some of which are likely to reach far into fundamental physics. Cast into questions, these problems can be enumerated as: (1) How can self-gravitating systems like dark-matter haloes form a universal density profile despite the absence of an equilibrium state? (2) Do the density profiles of galaxy clusters adopt the shape found in numerical simulations? (3) How concentrated is the dark matter in their cores? (4) Are galaxy clusters as substructured as expected for haloes formed from cold dark matter? (5) Are clusters as triaxial as they are expected to be?

Questions (1), (2) and possibly (3) address the fundamentally unsolved problem of what controls the non-equilibrium dynamics in of self-gravitating systems. Question (3) may be, question (4) certainly is related to the properties of the yet hypothetical, presumably weakly interacting, possibly annihilating dark-matter particle. Question (5) pertains our understanding of cosmic structure formation as being due to gravitational collapse from a Gaussian random field of density fluctuations.

\section{Baryonic physics in cluster cores}

In the centres of galaxy clusters, the mass fraction of baryons becomes significant. Whether it is large enough to substantially alter the properties of the dark matter ``backbone'' we do not know, but it is clear that one has to consider the influence of processes such as radiative cooling, heating and other feedback processes from baryons to dark matter when simulating cluster physics. Various processes act on scales that are orders of magnitude apart, which renders the simulation of the physics in cluster cores complicated. This is the reason why currently only approximate simulations of the conditions in cluster centres are available.

Similarly, the interpretation of observations is not easy. An outstanding example of the complications encountered in cluster centres is the Perseus cluster \citep[see][and references therein]{FA11.1}. This cluster can be studied in great detail due to its proximity and could thus be considered a prototype for the overall population. In this cluster, which on large scales is a regularly shaped, relaxed object, bubbles, ripples and shock fronts abound in the cluster core. There is a rich variety of phenomena including a complex gas temperature and metallicity structure, extended radio emission coming from the X-ray bubbles to filamentary structures that are bright in H-$\alpha$, but also visible in the X-ray regime. In this system, it is also apparent that projection effects have to be taken into account when trying to understand the complicated interactions.

It has been shown \citep[e.g.,][]{VO03.1} that the quantity $$K = \frac{T}{n_e^{3/2}}$$ has the properties of an entropy measure (``astronomer's entropy'') in the sense that it is constant during adiabatic processes. Here $T$ is the temperature and $n_e$ is the electron density, which is proportional to the gas density. When this quantity is plotted as a function of cluster radius and compared to results from non-radiative simulations \citep[e.g.,][]{PR10.1}, the deviations are strongest in the cluster core in the sense that the observed $K$ profiles lie above the numerical expectations. This is attributed to the effect of non-gravitational processes, or to processes that have acted before (``preheating'') or during the formation of the structures. It serves to show the influence of non-gravitational processes on cluster-sized structures. Already \citeauthor{PO03.2} (\citeyear{PO03.2}; see also the review by Giodini et al. in this volume) found that when computing the measure $K$ inside ten percent of the 
virial radius for several dozens of systems, their $K$-$T$ relation did not match the simple power-law $K\propto T$ expected for systems which share the same average gas density, for example if they formed at similar redshifts. They concluded that non-gravitational processes are in fact at work not only for the coolest of their studied systems, but also for systems with temperatures of 5~keV and above.

There is evidence for feedback processes in the centres of the most massive galaxy clusters. Here the gas densities of the order of $\sim 1\,\mathrm{particles\,cm^{-3}}$ {\em theoretically} imply large cooling rates of several hundreds of solar masses per year in some extreme cases \citep[e.g.][]{FA94.2, AL00.2}. Although steep profiles of the temperature as a function of radius (projected and de-projected) have been observed in many cases \citep[e.g.][]{AL01.1, GR02.2, VI05.1, PR07.1} that lead from $5$ to $10\,\mathrm{keV}$ at $r_{2500}$ for the most massive clusters down to around $1\,\mathrm{keV}$ in the cores of the clusters, no evidence has been found from radiative cooling that large quantities of gas are cooling below this value \citep[e.g.][]{PE03.1}. X-ray cavities reminiscent of those observed in the Perseus cluster have been found in many of these systems \citep[e.g.][]{AL06.1, BL11.1}. This suggests that heating through AGN feedback is an important mechanism in cluster cores. The exact way by 
which these different processes acting on very different size scales are balancing each other is still being studied \citep[e.g.][]{VA12.1}. Also, other processes such as thermal conduction \citep[e.g.][]{VO02.1} or mixing have been studied \citep{FA01.1} and are still under scrutiny.

A further process that has been repeatedly invoked in the interpretation of data is adiabatic contraction. The process leads to the contraction of the dark cluster halo due to the cooling and condensation of gas in cluster centres \citep[e.g.][]{GN04.1}. Although such models can be fitted to data, it is not clear yet whether this process is at work for galaxy clusters. Despite all the described ``baryonic impediments'', there is an intermediate region with radii of the order of the NFW scale radius where the effects of the central dark matter cusp are measurable, but where the feedback effects are not (yet) affecting the measurement significantly \citep[e.g.][]{PO05.1, SC07.4, VI06.1}. Here, X-ray emission from the baryonic component together with the assumption of hydrostatic equilibrium can be used to constrain the mass profile or the mass-concentration relation of galaxy clusters. The X-ray results on the cores of galaxies overall suggest a good agreement with the results from numerical simulations in 
terms of the mass profile \citep[e.g.,][and references therein]{AL01.1, AR02.1,LE03.1, VO06.1, SC07.4, HO11.2}. The determination of the amplitude of the mass-concentration relation is still somewhat uncertain and very likely affected by the large intrinsic scatter that is also observed in the simulations \citep{BU07.1, ET10.1, ET11.1, SC07.4}.

\section{Combining cluster observables}

How can a consistent picture of galaxy clusters be constructed that incorporates all kinds of information these objects provide? Of course, it is a viable approach to recover e.g.\ a cluster's radial mass profile by fitting a parametric model to X-ray and lensing data, as innumerable studies have shown. As the number of available empirical constraints on cluster mass distributions is increasing due to impressively improving observational data obtained in a multitude of channels, the question may be raised how all cluster observables may be combined on a common basis.

Current cluster observables are their gravitational lensing effects, their X-ray emission, the kinematics of their member galaxies, and the thermal Sunyaev-Zel'dovich effect. How can these different observables be combined on a common ground? The final section of this review sketches a possible approach based not on the mass or the mass density, but on the gravitational potential.

\subsection{Strong lensing and weak shear}

In the approximation relevant for our purposes, gravitational lensing is not affected by the physical state of the matter and energy distribution inside the lensing object. It thus provides perhaps the most reliable techniques for analysing the actual mass distribution in clusters \citep[see e.g.][for a review]{BA10.3}. Problems are generated by ambiguities due to \textit{a priori} unknown degrees of asymmetry, line-of-sight projection effects and the limited range of constraints in particular by strong gravitational lensing.

Strong lensing is being exposed in detail elsewhere in this volume. The main statements explained there are as follows: Strong lensing occurs with sources close to caustic curves, their images occur near critical curves. Two types of critical curve can be distinguished by the direction into which images are preferentially stretched. Tangential critical curves constrain the total projected amount of mass enclosed by them, albeit that mass is scaled by the critical surface mass density which depends on cosmology and sometimes unknown redshifts. Radial critical curves, which are typically surrounded by the tangential critical curves, constrain the local slope of the density profile.

The most prominent phenomena of strong gravitational lensing are tangential and radial arcs. The latter are difficult to find since they occur near cluster cores, where they may be overshone by the central cluster galaxies. Arcs trace or straddle the critical curves they are caused by. Multiple images, which are much more frequent than arcs, also constrain the locations of the critical curves.

One might think that strong lensing should constrain the core structure of galaxy clusters quite well. However, this is not the case if model-independent constraints are sought. Since all phenomena of strong lensing rely in some way or another on the occurrence and the locations of critical curves, all the information they provide concern the local curvature of the effective lensing potential. Strong-lensing observables can fix parametric mass models quite tightly, but they rarely allow model-independent statements on their own.

Weak gravitational lensing \citep[cf.][]{BA01.2} weakly distorts the images of background galaxies. To lowest order, weak lensing imprints an ellipticity on background sources characterised by the local strength of the gravitational tidal field or shear $\gamma$ and the scaled surface-mass density or convergence $\kappa$. Both $\kappa$ and $\gamma$ are linear combinations of the effective gravitational potential $\psi$. The shear has two components,
\begin{equation}
  \gamma_1 = \frac{1}{2}\left(\partial_1^2\psi-\partial_2^2\psi\right)\;,\quad
  \gamma_2 = \partial_1\partial_2\psi\;,
\label{eq:13}
\end{equation}
which can conveniently be combined to form the real and imaginary parts of the complex shear $\gamma = \gamma_1+\ii\gamma_2$. The convergence $\kappa$ is half the Laplacian of $\psi$,
\begin{equation}
  \kappa = \frac{1}{2}\left(\partial_1^2\psi+\partial_2^2\psi\right)\;.
\label{eq:14}
\end{equation}
To lowest-order approximation of weak lensing, idealised, circular sources appear as elliptical images. Let $a$ and $b$ be the semi-major and semi-minor axes of such an ellipse, then its ellipticity
\begin{equation}
  \epsilon = \frac{a-b}{a+b}
\label{eq:15}
\end{equation}
is approximately given by the so-called reduced shear $g$ rather than the shear itself,
\begin{equation}
  \epsilon \approx g := \frac{\gamma}{1-\kappa}\;.
\label{eq:16}
\end{equation}

Circular sources are rare if they exist at all, but can be approximately created by averaging over sufficiently many background sources if their intrinsic shapes are random and uncorrelated. Thus, weak-lensing effects have to be inferred in a statistical way by measuring the net ellipticity after averaging over images. This immediately implies that weak lensing has an intrinsic resolution limit set by the mean number density $n$ per solid angle of background sources. If $N$ galaxies should be averaged over, they need to be collected within a finite solid angle whose radius $\theta$ defines the angular resolution. The condition $N \gtrsim n\pi\theta^2$ requires
\begin{equation}
  \theta \gtrsim \sqrt{\frac{N}{n\pi}}\;.
\label{eq:17}
\end{equation}
Typical numbers are $n \approx (30\ldots40)\,\mathrm{arcmin^{-2}}$ and $N \approx 10$, giving $\theta \gtrsim (17\ldots20)''$. The resolution limit of weak-lensing observations thus turns out to be of the same order as the typical angular extent of the critical curves. Strong-lensing observables thus constrain effective cluster potentials on scales at or below the angular resolution provided by weak-lensing observables. If they are to be combined, and they need to be in order to constrain the core structure of galaxy clusters in a model-independent way, a multi-scale approach is required.

Such multi-scale approaches have been developed and applied \citep{BA96.3, BR06.1, BR05.2, CA06.1, ME10.2, ME09.2, ME11.1, UM12.1}. In essence, they map the effective lensing potential on a grid with a resolution adapted to the observed constraints, derive its lensing properties  by discrete differentiation and vary the gridded potential values such that a suitable total likelihood is minimised. This likelihood contains terms from strong and weak lensing. It expresses the expectation that the reconstructed lensing potential should recover the measured mean ellipticities and the location of the critical curves traced by any strong-lensing phenomena. Simulations show that such parameter-free, adaptive-grid methods recover the density profiles and the total cluster masses very well \citep{ME10.2}.

Multiple images constrain the deflection angle near critical curves, and thus the gradient of the lensing potential. As described elsewhere, critical curves are located where the determinant of the Jacobian matrix vanishes, which essentially provides a constraint on the matrix of second derivatives (the Hessian matrix) of the lensing potential. The gravitational shear measured from elliptical distortions constrains another combination of second-order derivatives of $\psi$. The different degrees of derivatives entering these constraints are conceptually important since the trace the lensing potential on different scales: The higher the derivative of the potential is that underlies an observable, the smaller are the scales that the respective observable is sensitive to. Aiming at the core structure and the amount of substructures in clusters, combining observables which are sensitive on different scales is conceptually important.

\subsection{Gravitational flexion}

Can third-order derivatives of the lensing potential be used? Their effect can straightforwardly be derived in the following way \citep{GO05.1, BA06.1}. Combine the derivatives with respect to the two spatial directions into one complex-valued differentiation operator $\partial$ and its complex conjugate $\partial^*$,
\begin{equation}
  \partial   := \partial_1+\ii\partial_2\;,\quad
  \partial^* := \partial_1-\ii\partial_2\;.
\label{eq:18}
\end{equation}
The convergence and the shear can then simply be written as
\begin{equation}
  \kappa = \frac{1}{2}\partial^*\partial\psi\;,\quad
  \gamma = \frac{1}{2}\partial^2\psi\;.
\label{eq:19}
\end{equation}
Further differentiation defines the two flexion components $F$ and $G$,
\begin{equation}
  F = \partial\kappa\;,\quad G = \partial\gamma\;,
\label{eq:20}
\end{equation}
with both $F$ and $G$ being complex numbers. Expressed in components, we have
\begin{equation}
  F_1 = \Re F = \frac{1}{2}\left(\psi_{111}+\psi_{122}\right)\;,\quad
  F_2 = \Im F = \frac{1}{2}\left(\psi_{112}+\psi_{222}\right)
\label{eq:21}
\end{equation}
and
\begin{equation}
  G_1 = \Re G = \frac{1}{2}\left(\psi_{111}-3\psi_{122}\right)\;,\quad
  G_2 = \Im G = \frac{1}{2}\left(3\psi_{112}-\psi_{222}\right)\;.
\label{eq:22}
\end{equation}
From the construction of the flexion components, it is quite easily seen that $F$ and $G$ create images with one- and three-fold rotational symmetry from circular sources, respectively. Such images are said to have spin-1 and spin-3. While the spin-1 image distortion due to the $F$ flexion causes a centroid shift which may be hard to recognise observationally, the spin-3 distortion due to the $G$ flexion imprints a characteristic triangular shape.

\begin{figure}[ht]
  \centering{\includegraphics[angle=90,width=0.75\hsize]{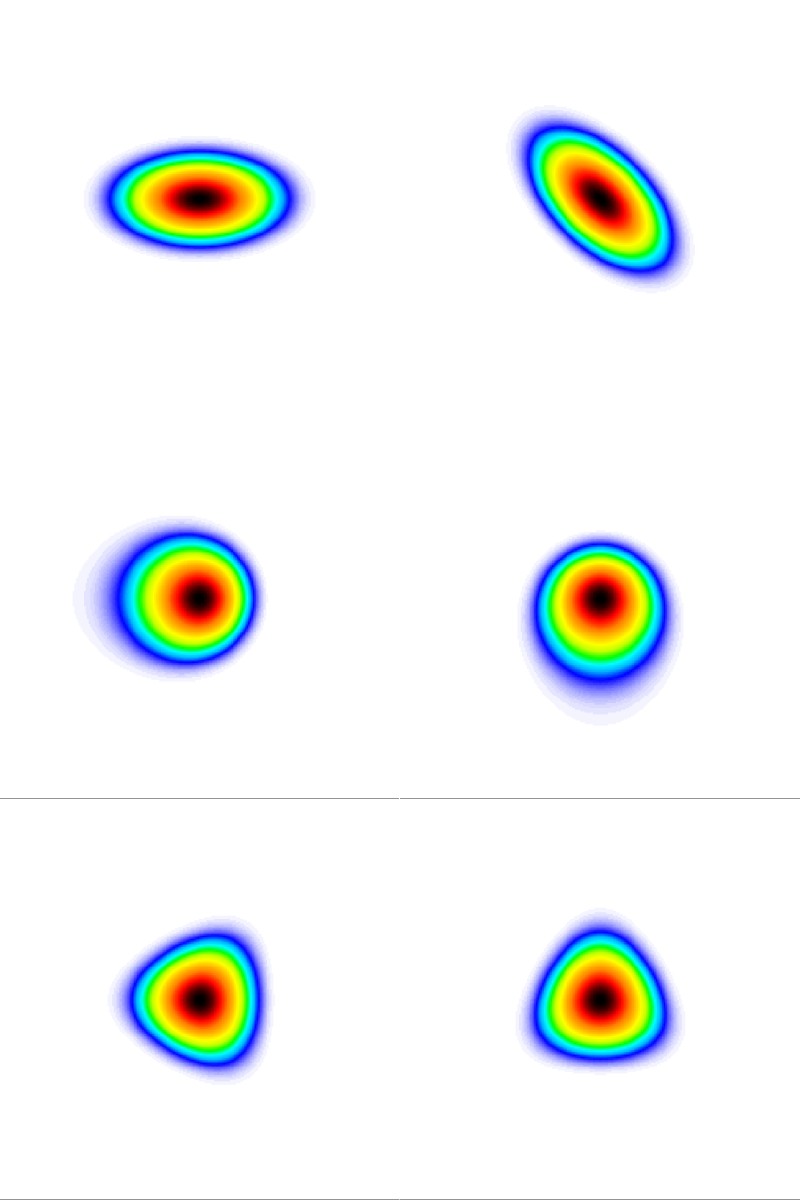}}
\caption{Illustration of gravitational lensing effects on a circular source. \textit{Left column}: shear; \textit{centre column}: $F$ flexion; \textit{right column}: $G$ flexion}
\label{fig:7}
\end{figure}

Faint galaxies tend to appear intrinsically elliptical. The gravitational shear, adding elliptical distortion, thus has to be detected on top of the intrinsic galaxy ellipticity. In contrast, since there should be little or no intrinsic triangular morphology in galaxies, the triangular distortion due to the $G$ flexion should be measurable with much lower intrinsic noise than the shear, and much less systematic uncertainty than the centroid shift caused by the $F$ flexion. The signal-to-noise ratio specifically of the $G$ flexion may thus be comparable to that of the shear, even though the signal itself is expected to be weaker.

To give an order of magnitude, consider the simple lens model of an isothermal sphere, whose intrinsic, three-dimensional density profile is proportional to $r^{-2}$. Its convergence and the absolute value of the shear are given by
\begin{equation}
  \kappa(\theta) = \frac{\theta_\mathrm{E}}{2\theta} = |\gamma(\theta)|\;,
\label{eq:23}
\end{equation}
where $\theta_\mathrm{E}$ is the Einstein radius. Clearly, $\kappa$ and $\gamma$ are dimension-less. The $F$ and $G$ flexions have the amplitude
\begin{equation}
  |F(\theta)| = \frac{\theta_\mathrm{E}}{2\theta^2} = |G(\theta)|
\label{eq:24}
\end{equation}
and the dimension of an inverse angle. Their signal, measured across an image of characteristic angular scale $\delta\theta$, is thus estimated to be
\begin{equation}
  |F(\theta)|\delta\theta = \frac{\theta_\mathrm{E}}{2\theta}\frac{\delta\theta}{\theta} =
  |G(\theta)|\delta\theta\;,
\label{eq:25}
\end{equation}
lower by a factor $\delta\theta/\theta$ than the shear signal. This factor is should fall between $0.1$ and $0.01$ for typical weakly-lensed images.

It is not clear yet whether it will be possible to reliably measure flexion in galaxy clusters. Some pioneering attempts have been made which found hints at $F$ rather than $G$ flexion \citep[e.g.][]{OK07.1, GO07.1, LE07.1, OK08.1, SC08.1, PI10.1, VE11.1, LE11.2, CA11.1}. However, flexion can only be measured if a reliable estimate for the shear already exists because any spin-1 or spin-3 quantity that can be derived from observations of galaxies not only picks up the flexion signal, but combinations of flexion and shear \citep{VI12.1}. Considerable work on shape measurements of galaxies and their interpretation still needs to be done before flexion can be considered measurable. If flexion can be determined, however, it promises to add small-scale information on the internal structure of galaxy clusters that the shear alone cannot resolve.

The essential message from this subsection is two-fold: Gravitational lensing primarily provides information on the second-order derivatives of the gravitational potential. This is necessarily so because of the equivalence principle: It is not the gravitational field that can be invariantly measured, but the gravitational tidal field, which is in the weak-field limit represented by the Hessian of the Newtonian potential. Should flexion become reliably measurable, it would allow quantifying changes in the tidal field, supplying smaller-scale information than the gravitational shear, the magnification and critical curves can provide. Lensing observables thus constrain the projected gravitational potential by its curvature and possibly higher-order derivatives. Gravitational lensing returns an estimate for the projected Newtonian potential up to harmonic functions.

\subsection{X-ray emission}

X-ray emission by thermal bremsstrahlung can also be analysed in terms of the gravitational potential. To illustrate this statement, assume that the X-ray emitting gas with density $\rho$, pressure $P$ and temperature $T$ is in hydrostatic equilibrium with the gravitational potential $\Phi$, satisfying
\begin{equation}
  \frac{\vec\nabla P}{\rho} = -\vec\nabla\Phi\;.
\label{eq:30}
\end{equation}
Assuming a polytropic stratification of an ideal gas with an effective polytropic index $\gamma$,
\begin{equation}
  P = P_0\left(\frac{\rho}{\rho_0}\right)^\gamma\;,
\label{eq:31}
\end{equation}
the hydrostatic equation (\ref{eq:30}) can easily be solved to give the gas density in terms of the gravitational potential,
\begin{equation}
  \rho = \rho_0\left[
    -\frac{\gamma-1}{\gamma}\frac{\rho_0}{P_0}\left(\Phi-\Phi_0\right)
  \right]^{1/(\gamma-1)}\;,
\label{eq:32}
\end{equation}
where the potential $\Phi_0$ enters as an integration constant that can be set to ensure $\rho = 0$ at a finite radius from the cluster core. By the ideal gas equation, (\ref{eq:32}) implies the relation
\begin{equation}
  T = T_0\left[
    -\frac{\gamma-1}{\gamma}\frac{\rho_0}{P_0}\left(\Phi-\Phi_0\right)
  \right]\;.
\label{eq:33}
\end{equation}
The fiducial pressure, density, and temperature, $P_0$, $\rho_0$ and $T_0$, can be fixed at an arbitrary cluster radius, e.g.\ at the centre. Equations (\ref{eq:32}) and (\ref{eq:33}) imply the (bolometric) bremsstrahlung emissivity
\begin{equation}
  j_\mathrm{X} = j_\mathrm{X, 0}\left[
    -\frac{\gamma-1}{\gamma}\frac{\rho_0}{P_0}\left(\Phi-\Phi_0\right)
  \right]^\alpha
\label{eq:34}
\end{equation}
with an exponent
\begin{equation}
  \alpha = \frac{2}{\gamma-1}+\frac{1}{2} = \frac{\gamma+3}{2(\gamma-1)}\;.
\label{eq:35}
\end{equation}
For $\gamma\approx1.2$, which may be appropriate for an estimate, $\alpha\approx10.5$: The X-ray emissivity depends very steeply on the gravitational potential. The exact value of the constant $j_\mathrm{X, 0}$ is known but irrelevant for our purposes.

The relation (\ref{eq:34}) that we have derived between the X-ray emissivity by thermal bremsstrahlung and the Newtonian potential suggests a path from the X-ray surface brightness $S_\mathrm{X}$ to the projected gravitational potential, thus connecting X-ray emission and gravitational lensing on a common ground. By means of Richardson-Lucy deprojection, an iterative scheme can be constructed allowing to recover the three-dimensional X-ray emissivity $j_\mathrm{X}$ from $S_\mathrm{X}$. The Richardson-Lucy deprojection rests on Bayes' theorem and returns, usually with a regularisation term suppressing noise overfitting, a three-dimensional function from its two-dimensional projection, provided a known projection kernel.

Notice that we have not applied any symmetry assumptions on the X-ray emissiting cluster so far. The integration of the hydrostatic equation leading to (\ref{eq:32}) and thus also to (\ref{eq:33}) is possible without adopting a particular symmetry. This has to be given up now before Richardson-Lucy deprojection can be applied because the projection kernel depends on the symmetry of the three-dimensional shape assumed for the body of the hot intracluster gas. Assuming a reasonable symmetry (which by no means has to be spherical), Richardson-Lucy deprojection returns an estimate for $j_\mathrm{X}$ from the observed $S_\mathrm{X}$.

Normalising the result, and raising it to the power of $1/\alpha$, (\ref{eq:34}) allows us to derive the scaled potential
\begin{equation}
  \frac{\rho_0}{P_0}\left(\Phi-\Phi_0\right) =
  -\frac{\gamma}{\gamma-1}\left(\frac{j_\mathrm{X}}{j_\mathrm{X, 0}}\right)^{1/\alpha}\;.
\label{eq:36}
\end{equation}
The exponent $\alpha$ needs to be known, but can be derived via the polytropic index $\gamma$ which can be determined from the temperature profile. Notice that the expression on the left-hand side of (\ref{eq:36}) is dimension-less since the potential is scaled by the squared thermal velocity $P_0/\rho_0$, which can again be estimated from the measured temperature and an assumed mean gas-particle mass. The scaled potential can then be projected along the line-of-sight to obtain an independent estimate of the projected Newtonian potential, which can directly be combined with the estimate of the projected potential $\psi$ provided by lensing. First tests of this algorithm assuming spherical symmetry returned very promising results.

\subsection{Thermal Sunyaev-Zel'dovich effect}

The same procedure developed above can now also be applied to observations of the thermal Sunyaev-Zel'dovich effect. This effect is quantified by the Compton-$y$ parameter
\begin{equation}
  y = \frac{k_\mathrm{B}T}{m_\mathrm{e}c^2}\sigma_\mathrm{T}n_\mathrm{e}\;,
\label{eq:37}
\end{equation}
where the relevant quantities are the electron number density $n_\mathrm{e}$ and the gas temperature $T$ and the other quantities are constants. Since the electron number density is determined by the gas density, we can use the results (\ref{eq:32}) and (\ref{eq:33}) obtained above to write
\begin{equation}
  y = y_0\left[
    -\frac{\gamma-1}{\gamma}\frac{\rho_0}{P_0}\left(\Phi-\Phi_0\right)
  \right]^\beta\;,
\label{eq:38}
\end{equation}
with the exponent
\begin{equation}
  \beta = \frac{1}{\gamma-1}+1 = \frac{\gamma}{\gamma-1} \approx 8.3\;.
\label{eq:39}
\end{equation}
Now, the same algorithm can be applied as for the X-ray surface brightness: The observed, line-of-sight integrated Compton-$y$ parameter $Y$ can be Richardson-Lucy deprojected to yield the three-dimensional Compton-$y$ parameter, which can the be converted to the (thermally scaled) gravitational potential and projected,
\begin{equation}
  \frac{\rho_0}{P_0}\left(\Phi-\Phi_0\right) =
  -\frac{\gamma}{\gamma-1}\left(\frac{y}{y_0}\right)^{1/\beta}\;.
\label{eq:40}
\end{equation}
Again, first tests show that this procedure returns reliable results.

The main result of the last two subsections is that the observables provided by the hot intracluster gas can directly be converted to constrain the projected Newtonian potential. This allows combining all related observables -- strong lensing, weak lensing possibly including flexion, X-ray emission and the thermal Sunyaev-Zel'dovich effect -- into one $\chi^2$ function returning a joint estimate for the projected cluster potential.

\subsection{Galaxy kinematics}

One observable remains, which is related to the kinematics of the cluster galaxies. Assuming spherical symmetry in the Jeans equation, we can write
\begin{equation}
  \frac{\partial(n\sigma_r^2)}{\partial r}+\frac{2\beta(r)}{r}(n\sigma_r^2) =
  -n\frac{\partial\Phi}{\partial r}\;,
\label{eq:41}
\end{equation} 
where $n$ is the spatial number density of the galaxies, $\sigma_r^2$ is their radial velocity dispersion, and
\begin{equation}
  \beta(r) = 1-\frac{\sigma_\theta^2}{\sigma_r^2}
\label{eq:42}
\end{equation}
is the usual anisotropy parameter of the velocity ellipsoid. We shall later assume for simplicity that the number density $n$ is proportional to the matter density $\rho$.

Instead of the usual approach, which would consist in solving (\ref{eq:41}) to arrive at a fairly involved integral relation between $(n\sigma_r^2)$ and $\partial_r\Phi$ for given $\beta(r)$, we can now follow a different route, relating the effective ``galaxy pressure'' $P_\mathrm{gal}=n\sigma_r^2$ directly to the matter density $\rho$ assuming a polytropic stratification with an effective polytropic index $\kappa$,
\begin{equation}
  P_\mathrm{gal} = P_\mathrm{gal, 0}\left(\frac{\rho}{\rho_0}\right)^\kappa =
  P_\mathrm{gal, 0}\left(\frac{n}{n_0}\right)^\kappa
\label{eq:43}
\end{equation}
Numerical experiments show that for a wide range of assumed matter-density profiles and anisotropy parameters, such effective polytropic indices can indeed be found. Then, (\ref{eq:41}) can be transformed into the equation
\begin{equation}
  \frac{\partial q}{\partial r}+\frac{2\epsilon\beta(r)}{r}q =
  -\frac{n_0}{P_0}\frac{\partial\Phi}{\partial r}
\label{eq:44}
\end{equation}
for the function
\begin{equation}
  q := \left(\frac{P_\mathrm{gal}}{P_\mathrm{gal, 0}}\right)^\epsilon
\label{eq:45}
\end{equation}
with the exponent
\begin{equation}
  \epsilon = \frac{\kappa-1}{\kappa}\;.
\label{eq:46}
\end{equation}
The equation (\ref{eq:44}) can now be solved in terms of $q$ first, resulting in a Volterra integral equation of the 2nd kind relating $q$ to the gravitational potential. Now, the line-of-sight integrated ``galaxy pressure'' is the observable line-of-sight velocity dispersion $\sigma_\parallel^2$, weighted by the galaxy number density. Again, by means of Richardson-Lucy deprojection, this observable can be converted to the three-dimensional ``galaxy pressure'' $P_\mathrm{gal}$, thus providing the measureable source term in the Volterra integral equation for the gravitational potential $\Phi$. This can then be solved by a standard, iterative procedure. The three-dimensional potential obtained this way can finally also be projected along the line-of-sight to return the projected gravitational potential $\psi$.

The main result of this subsection is that also the observable galaxy kinematics can be turned into a constraint on the two-dimensional gravitational potential, given an assumption on the anisotropy parameter of the galaxy orbits.

In conclusion, we have seen in the last section that all cluster observables can be combined on the grounds of the projected gravitational potential, jointly probing its structure on different scales with different resolution. First tests of these approaches have been undertaken with promising results. Details remain to be worked out, but it seems possible that this joint approach will return the tightest possible constraints on the gravitational potential, and thus on the density distribution, in clusters and their cores.

%
%
%
%
%
%
%


\begin{thebibliography}{119}
\ifx \bisbn   \undefined \def \bisbn  #1{ISBN #1}\fi
\ifx \binits  \undefined \def \binits#1{#1} \fi
\ifx \bauthor  \undefined \def \bauthor#1{#1} \fi
\ifx \bjtitle  \undefined \def \bjtitle#1{\textrm{#1}}\fi
\ifx \batitle  \undefined \def \batitle#1{#1} \fi
\ifx \bctitle  \undefined \def \bctitle#1{#1} \fi
\ifx \bvolume  \undefined \def \bvolume#1{\textbf{#1}}\fi
\ifx \byear  \undefined \def \byear#1{#1} \fi
\ifx \bissue  \undefined \def \bissue#1{#1} \fi
\ifx \bfpage  \undefined \def \bfpage#1{#1} \fi
\ifx \blpage  \undefined \def \blpage #1{#1} \fi
\ifx \burl  \undefined \def \burl#1{#1} \fi
\ifx \doiurl  \undefined \def \doiurl#1{#1} \fi
\ifx \betal  \undefined \def \betal{et al.} \fi
\ifx \binstitute  \undefined \def \binstitute#1{#1} \fi
\ifx \beditor  \undefined \def \beditor#1{#1} \fi
\ifx \bpublisher  \undefined \def \bpublisher#1{#1} \fi
\ifx \bbtitle  \undefined \def \bbtitle#1{\textit{#1}} \fi
\ifx \bedition  \undefined \def \bedition#1{#1} \fi
\ifx \bseriesno  \undefined \def \bseriesno#1{#1} \fi
\ifx \blocation  \undefined \def \blocation#1{#1} \fi
\ifx \bsertitle  \undefined \def \bsertitle#1{#1} \fi
\ifx \bsnm \undefined \def \bsnm#1{#1} \fi
\ifx \bsuffix \undefined \def \bsuffix#1{#1} \fi
\ifx \bparticle \undefined \def \bparticle#1{#1} \fi
\ifx \barticle \undefined \def \barticle#1{#1} \fi
\ifx \botherref \undefined \def \botherref #1{#1} \fi
\ifx \url \undefined \def \url#1{#1} \fi
\ifx \bchapter \undefined \def \bchapter#1{#1} \fi
\ifx \bbook \undefined \def \bbook#1{#1} \fi
\ifx \bcomment \undefined \def \bcomment#1{#1} \fi
\ifx \oauthor \undefined \def \oauthor#1{#1} \fi
\ifx \citeauthoryear \undefined \def \citeauthoryear#1{#1} \fi
\ifx \texttildelow  \undefined \def \texttildelow{\symbol{126}} \fi
\def \endbibitem {}
\ifx \bconflocation  \undefined \def \bconflocation#1{#1} \fi

\bibitem[\protect\citeauthoryear{{Abramowski} et~al.}{2012}]{AB12.1}
\begin{barticle}
\bauthor{\binits{A.} \bsnm{{Abramowski}}},
\bauthor{\binits{F.} \bsnm{{Acero}}},
\bauthor{\binits{F.} \bsnm{{Aharonian}}},
\bauthor{\binits{A.G.} \bsnm{{Akhperjanian}}},
\bauthor{\binits{G.} \bsnm{{Anton}}},
\bauthor{\binits{A.} \bsnm{{Balzer}}},
\bauthor{\binits{A.} \bsnm{{Barnacka}}},
\bauthor{\binits{U.} \bsnm{{Barres de Almeida}}},
\bauthor{\binits{Y.} \bsnm{{Becherini}}},
\bauthor{\binits{J.} \bsnm{{Becker}}},
\bauthor{\binits{B.} \bsnm{{Behera}}},
\bauthor{\binits{K.} \bsnm{{Bernl{\"o}hr}}},
\bauthor{\binits{E.} \bsnm{{Birsin}}},
\bauthor{\binits{J.} \bsnm{{Biteau}}},
\bauthor{\binits{A.} \bsnm{{Bochow}}},
\bauthor{\binits{C.} \bsnm{{Boisson}}},
\bauthor{\binits{J.} \bsnm{{Bolmont}}},
\bauthor{\binits{P.} \bsnm{{Bordas}}},
\bauthor{\binits{J.} \bsnm{{Brucker}}},
\bauthor{\binits{F.} \bsnm{{Brun}}},
\bauthor{\binits{P.} \bsnm{{Brun}}},
\bauthor{\binits{T.} \bsnm{{Bulik}}},
\bauthor{\binits{I.} \bsnm{{B{\"u}sching}}},
\bauthor{\binits{S.} \bsnm{{Carrigan}}},
\bauthor{\binits{S.} \bsnm{{Casanova}}},
\bauthor{\binits{M.} \bsnm{{Cerruti}}},
\bauthor{\binits{P.M.} \bsnm{{Chadwick}}},
\bauthor{\binits{A.} \bsnm{{Charbonnier}}},
\bauthor{\binits{R.C.G.} \bsnm{{Chaves}}},
\bauthor{\binits{A.} \bsnm{{Cheesebrough}}},
\bauthor{\binits{A.C.} \bsnm{{Clapson}}},
\bauthor{\binits{G.} \bsnm{{Coignet}}},
\bauthor{\binits{G.} \bsnm{{Cologna}}},
\bauthor{\binits{J.} \bsnm{{Conrad}}},
\bauthor{\binits{M.} \bsnm{{Dalton}}},
\bauthor{\binits{M.K.} \bsnm{{Daniel}}},
\bauthor{\binits{I.D.} \bsnm{{Davids}}},
\bauthor{\binits{B.} \bsnm{{Degrange}}},
\bauthor{\binits{C.} \bsnm{{Deil}}},
\bauthor{\binits{H.J.} \bsnm{{Dickinson}}},
\bauthor{\binits{A.} \bsnm{{Djannati-Ata{\"i}}}},
\bauthor{\binits{W.} \bsnm{{Domainko}}},
\bauthor{\binits{L.O.} \bsnm{{Drury}}},
\bauthor{\binits{G.} \bsnm{{Dubus}}},
\bauthor{\binits{K.} \bsnm{{Dutson}}},
\bauthor{\binits{J.} \bsnm{{Dyks}}},
\bauthor{\binits{M.} \bsnm{{Dyrda}}},
\bauthor{\binits{K.} \bsnm{{Egberts}}},
\bauthor{\binits{P.} \bsnm{{Eger}}},
\bauthor{\binits{P.} \bsnm{{Espigat}}},
\bauthor{\binits{L.} \bsnm{{Fallon}}},
\bauthor{\binits{C.} \bsnm{{Farnier}}},
\bauthor{\binits{S.} \bsnm{{Fegan}}},
\bauthor{\binits{F.} \bsnm{{Feinstein}}},
\bauthor{\binits{M.V.} \bsnm{{Fernandes}}},
\bauthor{\binits{A.} \bsnm{{Fiasson}}},
\bauthor{\binits{G.} \bsnm{{Fontaine}}},
\bauthor{\binits{A.} \bsnm{{F{\"o}rster}}},
\bauthor{\binits{M.} \bsnm{{F{\"u}{\ss}ling}}},
\bauthor{\binits{Y.A.} \bsnm{{Gallant}}},
\bauthor{\binits{H.} \bsnm{{Gast}}},
\bauthor{\binits{L.} \bsnm{{G{\'e}rard}}},
\bauthor{\binits{D.} \bsnm{{Gerbig}}},
\bauthor{\binits{B.} \bsnm{{Giebels}}},
\bauthor{\binits{J.F.} \bsnm{{Glicenstein}}},
\bauthor{\binits{B.} \bsnm{{Gl{\"u}ck}}},
\bauthor{\binits{P.} \bsnm{{Goret}}},
\bauthor{\binits{D.} \bsnm{{G{\"o}ring}}},
\bauthor{\binits{S.} \bsnm{{H{\"a}ffner}}},
\bauthor{\binits{J.D.} \bsnm{{Hague}}},
\bauthor{\binits{D.} \bsnm{{Hampf}}},
\bauthor{\binits{M.} \bsnm{{Hauser}}},
\bauthor{\binits{S.} \bsnm{{Heinz}}},
\bauthor{\binits{G.} \bsnm{{Heinzelmann}}},
\bauthor{\binits{G.} \bsnm{{Henri}}},
\bauthor{\binits{G.} \bsnm{{Hermann}}},
\bauthor{\binits{J.A.} \bsnm{{Hinton}}},
\bauthor{\binits{A.} \bsnm{{Hoffmann}}},
\bauthor{\binits{W.} \bsnm{{Hofmann}}},
\bauthor{\binits{P.} \bsnm{{Hofverberg}}},
\bauthor{\binits{M.} \bsnm{{Holler}}},
\bauthor{\binits{D.} \bsnm{{Horns}}},
\bauthor{\binits{A.} \bsnm{{Jacholkowska}}},
\bauthor{\binits{O.C.} \bsnm{{de Jager}}},
\bauthor{\binits{C.} \bsnm{{Jahn}}},
\bauthor{\binits{M.} \bsnm{{Jamrozy}}},
\bauthor{\binits{I.} \bsnm{{Jung}}},
\bauthor{\binits{M.A.} \bsnm{{Kastendieck}}},
\bauthor{\binits{K.} \bsnm{{Katarzy{\'n}ski}}},
\bauthor{\binits{U.} \bsnm{{Katz}}},
\bauthor{\binits{S.} \bsnm{{Kaufmann}}},
\bauthor{\binits{D.} \bsnm{{Keogh}}},
\bauthor{\binits{D.} \bsnm{{Khangulyan}}},
\bauthor{\binits{B.} \bsnm{{Kh{\'e}lifi}}},
\bauthor{\binits{D.} \bsnm{{Klochkov}}},
\bauthor{\binits{W.} \bsnm{{Klu{\'z}niak}}},
\bauthor{\binits{T.} \bsnm{{Kneiske}}},
\bauthor{\binits{N.} \bsnm{{Komin}}},
\bauthor{\binits{K.} \bsnm{{Kosack}}},
\bauthor{\binits{R.} \bsnm{{Kossakowski}}},
\bauthor{\binits{H.} \bsnm{{Laffon}}},
\bauthor{\binits{G.} \bsnm{{Lamanna}}},
\bauthor{\binits{D.} \bsnm{{Lennarz}}},
\bauthor{\binits{T.} \bsnm{{Lohse}}},
\bauthor{\binits{A.} \bsnm{{Lopatin}}},
\bauthor{\binits{C.-C.} \bsnm{{Lu}}},
\bauthor{\binits{V.} \bsnm{{Marandon}}},
\bauthor{\binits{A.} \bsnm{{Marcowith}}},
\bauthor{\binits{J.} \bsnm{{Masbou}}},
\bauthor{\binits{D.} \bsnm{{Maurin}}},
\bauthor{\binits{N.} \bsnm{{Maxted}}},
\bauthor{\binits{M.} \bsnm{{Mayer}}},
\bauthor{\binits{T.J.L.} \bsnm{{McComb}}},
\bauthor{\binits{M.C.} \bsnm{{Medina}}},
\bauthor{\binits{J.} \bsnm{{M{\'e}hault}}},
\bauthor{\binits{R.} \bsnm{{Moderski}}},
\bauthor{\binits{E.} \bsnm{{Moulin}}},
\bauthor{\binits{C.L.} \bsnm{{Naumann}}},
\bauthor{\binits{M.} \bsnm{{Naumann-Godo}}},
\bauthor{\binits{M.} \bsnm{{de Naurois}}},
\bauthor{\binits{D.} \bsnm{{Nedbal}}},
\bauthor{\binits{D.} \bsnm{{Nekrassov}}},
\bauthor{\binits{N.} \bsnm{{Nguyen}}},
\bauthor{\binits{B.} \bsnm{{Nicholas}}},
\bauthor{\binits{J.} \bsnm{{Niemiec}}},
\bauthor{\binits{S.J.} \bsnm{{Nolan}}},
\bauthor{\binits{S.} \bsnm{{Ohm}}},
\bauthor{\binits{E.} \bsnm{{de O{\~n}a Wilhelmi}}},
\bauthor{\binits{B.} \bsnm{{Opitz}}},
\bauthor{\binits{M.} \bsnm{{Ostrowski}}},
\bauthor{\binits{I.} \bsnm{{Oya}}},
\bauthor{\binits{M.} \bsnm{{Panter}}},
\bauthor{\binits{M.} \bsnm{{Paz Arribas}}},
\bauthor{\binits{G.} \bsnm{{Pedaletti}}},
\bauthor{\binits{G.} \bsnm{{Pelletier}}},
\bauthor{\binits{P.-O.} \bsnm{{Petrucci}}},
\bauthor{\binits{S.} \bsnm{{Pita}}},
\bauthor{\binits{G.} \bsnm{{P{\"u}hlhofer}}},
\bauthor{\binits{M.} \bsnm{{Punch}}},
\bauthor{\binits{A.} \bsnm{{Quirrenbach}}},
\bauthor{\binits{M.} \bsnm{{Raue}}},
\bauthor{\binits{S.M.} \bsnm{{Rayner}}},
\bauthor{\binits{A.} \bsnm{{Reimer}}},
\bauthor{\binits{O.} \bsnm{{Reimer}}},
\bauthor{\binits{M.} \bsnm{{Renaud}}},
\bauthor{\binits{R.} \bsnm{{de los Reyes}}},
\bauthor{\binits{F.} \bsnm{{Rieger}}},
\bauthor{\binits{J.} \bsnm{{Ripken}}},
\bauthor{\binits{L.} \bsnm{{Rob}}},
\bauthor{\binits{S.} \bsnm{{Rosier-Lees}}},
\bauthor{\binits{G.} \bsnm{{Rowell}}},
\bauthor{\binits{B.} \bsnm{{Rudak}}},
\bauthor{\binits{C.B.} \bsnm{{Rulten}}},
\bauthor{\binits{J.} \bsnm{{Ruppel}}},
\bauthor{\binits{V.} \bsnm{{Sahakian}}},
\bauthor{\binits{D.A.} \bsnm{{Sanchez}}},
\bauthor{\binits{A.} \bsnm{{Santangelo}}},
\bauthor{\binits{R.} \bsnm{{Schlickeiser}}},
\bauthor{\binits{F.M.} \bsnm{{Sch{\"o}ck}}},
\bauthor{\binits{A.} \bsnm{{Schulz}}},
\bauthor{\binits{U.} \bsnm{{Schwanke}}},
\bauthor{\binits{S.} \bsnm{{Schwarzburg}}},
\bauthor{\binits{S.} \bsnm{{Schwemmer}}},
\bauthor{\binits{F.} \bsnm{{Sheidaei}}},
\bauthor{\binits{J.L.} \bsnm{{Skilton}}},
\bauthor{\binits{H.} \bsnm{{Sol}}},
\bauthor{\binits{G.} \bsnm{{Spengler}}},
\bauthor{\binits{{\L}.} \bsnm{{Stawarz}}},
\bauthor{\binits{R.} \bsnm{{Steenkamp}}},
\bauthor{\binits{C.} \bsnm{{Stegmann}}},
\bauthor{\binits{F.} \bsnm{{Stinzing}}},
\bauthor{\binits{K.} \bsnm{{Stycz}}},
\bauthor{\binits{I.} \bsnm{{Sushch}}},
\bauthor{\binits{A.} \bsnm{{Szostek}}},
\bauthor{\binits{J.-P.} \bsnm{{Tavernet}}},
\bauthor{\binits{R.} \bsnm{{Terrier}}},
\bauthor{\binits{M.} \bsnm{{Tluczykont}}},
\bauthor{\binits{K.} \bsnm{{Valerius}}},
\bauthor{\binits{C.} \bsnm{{van Eldik}}},
\bauthor{\binits{G.} \bsnm{{Vasileiadis}}},
\bauthor{\binits{C.} \bsnm{{Venter}}},
\bauthor{\binits{J.P.} \bsnm{{Vialle}}},
\bauthor{\binits{A.} \bsnm{{Viana}}},
\bauthor{\binits{P.} \bsnm{{Vincent}}},
\bauthor{\binits{H.J.} \bsnm{{V{\"o}lk}}},
\bauthor{\binits{F.} \bsnm{{Volpe}}},
\bauthor{\binits{S.} \bsnm{{Vorobiov}}},
\bauthor{\binits{M.} \bsnm{{Vorster}}},
\bauthor{\binits{S.J.} \bsnm{{Wagner}}},
\bauthor{\binits{M.} \bsnm{{Ward}}},
\bauthor{\binits{R.} \bsnm{{White}}},
\bauthor{\binits{A.} \bsnm{{Wierzcholska}}},
\bauthor{\binits{M.} \bsnm{{Zacharias}}},
\bauthor{\binits{A.} \bsnm{{Zajczyk}}},
\bauthor{\binits{A.A.} \bsnm{{Zdziarski}}},
\bauthor{\binits{A.} \bsnm{{Zech}}},
\bauthor{\binits{H.-S.} \bsnm{{Zechlin}}},
\bauthor{\bsnm{{H.~E.~S.~S.~Collaboration}}},
\batitle{Search for dark matter annihilation signals from the fornax galaxy
  cluster with h.e.s.s.}
\bjtitle{ApJ}
\bvolume{750},
\bfpage{123}
(\byear{2012}).
doi:\doiurl{10.1088/0004-637X/750/2/123}
\end{barticle}
\endbibitem

\bibitem[\protect\citeauthoryear{{Allen}}{2000}]{AL00.2}
\begin{barticle}
\bauthor{\binits{S.W.} \bsnm{{Allen}}},
\batitle{The properties of cooling flows in x-ray luminous clusters of
  galaxies}.
\bjtitle{MNRAS}
\bvolume{315},
\bfpage{269}--\blpage{295}
(\byear{2000}).
doi:\doiurl{10.1046/j.1365-8711.2000.03395.x}
\end{barticle}
\endbibitem

\bibitem[\protect\citeauthoryear{{Allen} et~al.}{2001}]{AL01.1}
\begin{barticle}
\bauthor{\binits{S.W.} \bsnm{{Allen}}},
\bauthor{\binits{S.} \bsnm{{Ettori}}},
\bauthor{\binits{A.C.} \bsnm{{Fabian}}},
\batitle{Chandra measurements of the distribution of mass in the luminous
  lensing cluster abell 2390}.
\bjtitle{MNRAS}
\bvolume{324},
\bfpage{877}--\blpage{890}
(\byear{2001}).
doi:\doiurl{10.1046/j.1365-8711.2001.04318.x}
\end{barticle}
\endbibitem

\bibitem[\protect\citeauthoryear{{Allen} et~al.}{2006}]{AL06.1}
\begin{barticle}
\bauthor{\binits{S.W.} \bsnm{{Allen}}},
\bauthor{\binits{R.J.H.} \bsnm{{Dunn}}},
\bauthor{\binits{A.C.} \bsnm{{Fabian}}},
\bauthor{\binits{G.B.} \bsnm{{Taylor}}},
\bauthor{\binits{C.S.} \bsnm{{Reynolds}}},
\batitle{The relation between accretion rate and jet power in x-ray luminous
  elliptical galaxies}.
\bjtitle{MNRAS}
\bvolume{372},
\bfpage{21}--\blpage{30}
(\byear{2006}).
doi:\doiurl{10.1111/j.1365-2966.2006.10778.x}
\end{barticle}
\endbibitem

\bibitem[\protect\citeauthoryear{{Ando} and {Nagai}}{2012}]{AN12.1}
\begin{barticle}
\bauthor{\binits{S.} \bsnm{{Ando}}},
\bauthor{\binits{D.} \bsnm{{Nagai}}},
\batitle{Fermi-lat constraints on dark matter annihilation cross section from
  observations of the fornax cluster}.
\bjtitle{JCAP}
\bvolume{7},
\bfpage{17}
(\byear{2012}).
doi:\doiurl{10.1088/1475-7516/2012/07/017}
\end{barticle}
\endbibitem

\bibitem[\protect\citeauthoryear{{Arabadjis} et~al.}{2002}]{AR02.1}
\begin{barticle}
\bauthor{\binits{J.S.} \bsnm{{Arabadjis}}},
\bauthor{\binits{M.W.} \bsnm{{Bautz}}},
\bauthor{\binits{G.P.} \bsnm{{Garmire}}},
\batitle{Chandra observations of the lensing cluster emss 1358+6245:
  Implications for self-interacting dark matter}.
\bjtitle{ApJ}
\bvolume{572},
\bfpage{66}--\blpage{78}
(\byear{2002})
\end{barticle}
\endbibitem

\bibitem[\protect\citeauthoryear{{Ascasibar} et~al.}{2007}]{AS07.1}
\begin{barticle}
\bauthor{\binits{Y.} \bsnm{{Ascasibar}}},
\bauthor{\binits{Y.} \bsnm{{Hoffman}}},
\bauthor{\binits{S.} \bsnm{{Gottl{\"o}ber}}},
\batitle{Secondary infall and dark matter haloes}.
\bjtitle{MNRAS}
\bvolume{376},
\bfpage{393}--\blpage{404}
(\byear{2007}).
doi:\doiurl{10.1111/j.1365-2966.2007.11439.x}
\end{barticle}
\endbibitem

\bibitem[\protect\citeauthoryear{{Austin} et~al.}{2005}]{AU05.1}
\begin{barticle}
\bauthor{\binits{C.G.} \bsnm{{Austin}}},
\bauthor{\binits{L.L.R.} \bsnm{{Williams}}},
\bauthor{\binits{E.I.} \bsnm{{Barnes}}},
\bauthor{\binits{A.} \bsnm{{Babul}}},
\bauthor{\binits{J.J.} \bsnm{{Dalcanton}}},
\batitle{Semianalytical dark matter halos and the jeans equation}.
\bjtitle{ApJ}
\bvolume{634},
\bfpage{756}--\blpage{774}
(\byear{2005}).
doi:\doiurl{10.1086/497133}
\end{barticle}
\endbibitem

\bibitem[\protect\citeauthoryear{{Bacon} et~al.}{2006}]{BA06.1}
\begin{barticle}
\bauthor{\binits{D.J.} \bsnm{{Bacon}}},
\bauthor{\binits{D.M.} \bsnm{{Goldberg}}},
\bauthor{\binits{B.T.P.} \bsnm{{Rowe}}},
\bauthor{\binits{A.N.} \bsnm{{Taylor}}},
\batitle{Weak gravitational flexion}.
\bjtitle{MNRAS}
\bvolume{365},
\bfpage{414}--\blpage{428}
(\byear{2006}).
doi:\doiurl{10.1111/j.1365-2966.2005.09624.x}
\end{barticle}
\endbibitem

\bibitem[\protect\citeauthoryear{{Bah{\'e}} et~al.}{2012}]{BA12.1}
\begin{barticle}
\bauthor{\binits{Y.M.} \bsnm{{Bah{\'e}}}},
\bauthor{\binits{I.G.} \bsnm{{McCarthy}}},
\bauthor{\binits{L.J.} \bsnm{{King}}},
\batitle{Mock weak lensing analysis of simulated galaxy clusters: bias and
  scatter in mass and concentration}.
\bjtitle{MNRAS}
\bvolume{421},
\bfpage{1073}--\blpage{1088}
(\byear{2012}).
doi:\doiurl{10.1111/j.1365-2966.2011.20364.x}
\end{barticle}
\endbibitem

\bibitem[\protect\citeauthoryear{{Bartelmann}}{2010}]{BA10.3}
\begin{barticle}
\bauthor{\binits{M.} \bsnm{{Bartelmann}}},
\batitle{Topical review gravitational lensing}.
\bjtitle{Classical and Quantum Gravity}
\bvolume{27}(\bissue{23}),
\bfpage{233001}
(\byear{2010}).
doi:\doiurl{10.1088/0264-9381/27/23/233001}
\end{barticle}
\endbibitem

\bibitem[\protect\citeauthoryear{{Bartelmann} and {Schneider}}{2001}]{BA01.2}
\begin{barticle}
\bauthor{\binits{M.} \bsnm{{Bartelmann}}},
\bauthor{\binits{P.} \bsnm{{Schneider}}},
\batitle{Weak gravitational lensing}.
\bjtitle{Phys. Rep.}
\bvolume{340},
\bfpage{291}--\blpage{472}
(\byear{2001})
\end{barticle}
\endbibitem

\bibitem[\protect\citeauthoryear{{Bartelmann} et~al.}{1996}]{BA96.3}
\begin{barticle}
\bauthor{\binits{M.} \bsnm{{Bartelmann}}},
\bauthor{\binits{R.} \bsnm{{Narayan}}},
\bauthor{\binits{S.} \bsnm{{Seitz}}},
\bauthor{\binits{P.} \bsnm{{Schneider}}},
\batitle{Maximum-likelihood cluster reconstruction}.
\bjtitle{ApJL}
\bvolume{464},
\bfpage{115}
(\byear{1996})
\end{barticle}
\endbibitem

\bibitem[\protect\citeauthoryear{{Blanton} et~al.}{2011}]{BL11.1}
\begin{barticle}
\bauthor{\binits{E.L.} \bsnm{{Blanton}}},
\bauthor{\binits{S.W.} \bsnm{{Randall}}},
\bauthor{\binits{T.E.} \bsnm{{Clarke}}},
\bauthor{\binits{C.L.} \bsnm{{Sarazin}}},
\bauthor{\binits{B.R.} \bsnm{{McNamara}}},
\bauthor{\binits{E.M.} \bsnm{{Douglass}}},
\bauthor{\binits{M.} \bsnm{{McDonald}}},
\batitle{A very deep chandra observation of a2052: Bubbles, shocks, and
  sloshing}.
\bjtitle{ApJ}
\bvolume{737},
\bfpage{99}
(\byear{2011}).
doi:\doiurl{10.1088/0004-637X/737/2/99}
\end{barticle}
\endbibitem

\bibitem[\protect\citeauthoryear{{Boylan-Kolchin} et~al.}{2009}]{BO09.1}
\begin{barticle}
\bauthor{\binits{M.} \bsnm{{Boylan-Kolchin}}},
\bauthor{\binits{V.} \bsnm{{Springel}}},
\bauthor{\binits{S.D.M.} \bsnm{{White}}},
\bauthor{\binits{A.} \bsnm{{Jenkins}}},
\bauthor{\binits{G.} \bsnm{{Lemson}}},
\batitle{Resolving cosmic structure formation with the millennium-ii
  simulation}.
\bjtitle{MNRAS}
\bvolume{398},
\bfpage{1150}--\blpage{1164}
(\byear{2009}).
doi:\doiurl{10.1111/j.1365-2966.2009.15191.x}
\end{barticle}
\endbibitem

\bibitem[\protect\citeauthoryear{{Brada{\v c}} et~al.}{2005}]{BR05.2}
\begin{barticle}
\bauthor{\binits{M.} \bsnm{{Brada{\v c}}}},
\bauthor{\binits{T.} \bsnm{{Erben}}},
\bauthor{\binits{P.} \bsnm{{Schneider}}},
\bauthor{\binits{H.} \bsnm{{Hildebrandt}}},
\bauthor{\binits{M.} \bsnm{{Lombardi}}},
\bauthor{\binits{M.} \bsnm{{Schirmer}}},
\bauthor{\binits{J.} \bsnm{{Miralles}}},
\bauthor{\binits{D.} \bsnm{{Clowe}}},
\bauthor{\binits{S.} \bsnm{{Schindler}}},
\batitle{Strong and weak lensing united}.
\bjtitle{A\&A}
\bvolume{437},
\bfpage{49}--\blpage{60}
(\byear{2005}).
doi:\doiurl{10.1051/0004-6361:20042234}
\end{barticle}
\endbibitem

\bibitem[\protect\citeauthoryear{{Brada{\v c}} et~al.}{2006}]{BR06.1}
\begin{barticle}
\bauthor{\binits{M.} \bsnm{{Brada{\v c}}}},
\bauthor{\binits{D.} \bsnm{{Clowe}}},
\bauthor{\binits{A.H.} \bsnm{{Gonzalez}}},
\bauthor{\binits{P.} \bsnm{{Marshall}}},
\bauthor{\binits{W.} \bsnm{{Forman}}},
\bauthor{\binits{C.} \bsnm{{Jones}}},
\bauthor{\binits{M.} \bsnm{{Markevitch}}},
\bauthor{\binits{S.} \bsnm{{Randall}}},
\bauthor{\binits{T.} \bsnm{{Schrabback}}},
\bauthor{\binits{D.} \bsnm{{Zaritsky}}},
\batitle{Strong and weak lensing united. iii. measuring the mass distribution
  of the merging galaxy cluster 1es 0657-558}.
\bjtitle{ApJ}
\bvolume{652},
\bfpage{937}--\blpage{947}
(\byear{2006}).
doi:\doiurl{10.1086/508601}
\end{barticle}
\endbibitem

\bibitem[\protect\citeauthoryear{{Broadhurst} et~al.}{2005}]{BR05.1}
\begin{barticle}
\bauthor{\binits{T.} \bsnm{{Broadhurst}}},
\bauthor{\binits{N.} \bsnm{{Ben{\'{\i}}tez}}},
\bauthor{\binits{D.} \bsnm{{Coe}}},
\bauthor{\binits{K.} \bsnm{{Sharon}}},
\bauthor{\binits{K.} \bsnm{{Zekser}}},
\bauthor{\binits{R.} \bsnm{{White}}},
\bauthor{\binits{H.} \bsnm{{Ford}}},
\bauthor{\binits{R.} \bsnm{{Bouwens}}},
\bauthor{\binits{J.} \bsnm{{Blakeslee}}},
\bauthor{\binits{M.} \bsnm{{Clampin}}},
\bauthor{\binits{N.} \bsnm{{Cross}}},
\bauthor{\binits{M.} \bsnm{{Franx}}},
\bauthor{\binits{B.} \bsnm{{Frye}}},
\bauthor{\binits{G.} \bsnm{{Hartig}}},
\bauthor{\binits{G.} \bsnm{{Illingworth}}},
\bauthor{\binits{L.} \bsnm{{Infante}}},
\bauthor{\binits{F.} \bsnm{{Menanteau}}},
\bauthor{\binits{G.} \bsnm{{Meurer}}},
\bauthor{\binits{M.} \bsnm{{Postman}}},
\bauthor{\binits{D.R.} \bsnm{{Ardila}}},
\bauthor{\binits{F.} \bsnm{{Bartko}}},
\bauthor{\binits{R.A.} \bsnm{{Brown}}},
\bauthor{\binits{C.J.} \bsnm{{Burrows}}},
\bauthor{\binits{E.S.} \bsnm{{Cheng}}},
\bauthor{\binits{P.D.} \bsnm{{Feldman}}},
\bauthor{\binits{D.A.} \bsnm{{Golimowski}}},
\bauthor{\binits{T.} \bsnm{{Goto}}},
\bauthor{\binits{C.} \bsnm{{Gronwall}}},
\bauthor{\binits{D.} \bsnm{{Herranz}}},
\bauthor{\binits{B.} \bsnm{{Holden}}},
\bauthor{\binits{N.} \bsnm{{Homeier}}},
\bauthor{\binits{J.E.} \bsnm{{Krist}}},
\bauthor{\binits{M.P.} \bsnm{{Lesser}}},
\bauthor{\binits{A.R.} \bsnm{{Martel}}},
\bauthor{\binits{G.K.} \bsnm{{Miley}}},
\bauthor{\binits{P.} \bsnm{{Rosati}}},
\bauthor{\binits{M.} \bsnm{{Sirianni}}},
\bauthor{\binits{W.B.} \bsnm{{Sparks}}},
\bauthor{\binits{S.} \bsnm{{Steindling}}},
\bauthor{\binits{H.D.} \bsnm{{Tran}}},
\bauthor{\binits{Z.I.} \bsnm{{Tsvetanov}}},
\bauthor{\binits{W.} \bsnm{{Zheng}}},
\batitle{Strong-lensing analysis of a1689 from deep advanced camera images}.
\bjtitle{ApJ}
\bvolume{621},
\bfpage{53}--\blpage{88}
(\byear{2005}).
doi:\doiurl{10.1086/426494}
\end{barticle}
\endbibitem

\bibitem[\protect\citeauthoryear{{Broadhurst} et~al.}{2008}]{BR08.3}
\begin{barticle}
\bauthor{\binits{T.} \bsnm{{Broadhurst}}},
\bauthor{\binits{K.} \bsnm{{Umetsu}}},
\bauthor{\binits{E.} \bsnm{{Medezinski}}},
\bauthor{\binits{M.} \bsnm{{Oguri}}},
\bauthor{\binits{Y.} \bsnm{{Rephaeli}}},
\batitle{Comparison of cluster lensing profiles with {$\Lambda$}cdm
  predictions}.
\bjtitle{ApJL}
\bvolume{685},
\bfpage{9}--\blpage{12}
(\byear{2008}).
doi:\doiurl{10.1086/592400}
\end{barticle}
\endbibitem

\bibitem[\protect\citeauthoryear{{Bullock} et~al.}{2001}]{BU01.1}
\begin{barticle}
\bauthor{\binits{J.S.} \bsnm{{Bullock}}},
\bauthor{\binits{T.S.} \bsnm{{Kolatt}}},
\bauthor{\binits{Y.} \bsnm{{Sigad}}},
\bauthor{\binits{R.S.} \bsnm{{Somerville}}},
\bauthor{\binits{A.V.} \bsnm{{Kravtsov}}},
\bauthor{\binits{A.A.} \bsnm{{Klypin}}},
\bauthor{\binits{J.R.} \bsnm{{Primack}}},
\bauthor{\binits{A.} \bsnm{{Dekel}}},
\batitle{Profiles of dark haloes: evolution, scatter and environment}.
\bjtitle{MNRAS}
\bvolume{321},
\bfpage{559}--\blpage{575}
(\byear{2001}).
doi:\doiurl{10.1046/j.1365-8711.2001.04068.x}
\end{barticle}
\endbibitem

\bibitem[\protect\citeauthoryear{{Buote} and {Lewis}}{2004}]{BU04.1}
\begin{barticle}
\bauthor{\binits{D.A.} \bsnm{{Buote}}},
\bauthor{\binits{A.D.} \bsnm{{Lewis}}},
\batitle{The dark matter radial profile in the core of the relaxed cluster
  a2589}.
\bjtitle{ApJ}
\bvolume{604},
\bfpage{116}--\blpage{124}
(\byear{2004}).
doi:\doiurl{10.1086/381793}
\end{barticle}
\endbibitem

\bibitem[\protect\citeauthoryear{{Buote} et~al.}{2007}]{BU07.1}
\begin{barticle}
\bauthor{\binits{D.A.} \bsnm{{Buote}}},
\bauthor{\binits{F.} \bsnm{{Gastaldello}}},
\bauthor{\binits{P.J.} \bsnm{{Humphrey}}},
\bauthor{\binits{L.} \bsnm{{Zappacosta}}},
\bauthor{\binits{J.S.} \bsnm{{Bullock}}},
\bauthor{\binits{F.} \bsnm{{Brighenti}}},
\bauthor{\binits{W.G.} \bsnm{{Mathews}}},
\batitle{The x-ray concentration-virial mass relation}.
\bjtitle{ApJ}
\bvolume{664},
\bfpage{123}--\blpage{134}
(\byear{2007}).
doi:\doiurl{10.1086/518684}
\end{barticle}
\endbibitem

\bibitem[\protect\citeauthoryear{{Cacciato} et~al.}{2006}]{CA06.1}
\begin{barticle}
\bauthor{\binits{M.} \bsnm{{Cacciato}}},
\bauthor{\binits{M.} \bsnm{{Bartelmann}}},
\bauthor{\binits{M.} \bsnm{{Meneghetti}}},
\bauthor{\binits{L.} \bsnm{{Moscardini}}},
\batitle{Combining weak and strong lensing in cluster potential
  reconstruction}.
\bjtitle{A\&A}
\bvolume{458},
\bfpage{349}--\blpage{356}
(\byear{2006}).
doi:\doiurl{10.1051/0004-6361:20054582}
\end{barticle}
\endbibitem

\bibitem[\protect\citeauthoryear{{Cain} et~al.}{2011}]{CA11.1}
\begin{barticle}
\bauthor{\binits{B.} \bsnm{{Cain}}},
\bauthor{\binits{P.L.} \bsnm{{Schechter}}},
\bauthor{\binits{M.W.} \bsnm{{Bautz}}},
\batitle{Measuring gravitational lensing flexion in a1689 using an analytic
  image model}.
\bjtitle{ApJ}
\bvolume{736},
\bfpage{43}
(\byear{2011}).
doi:\doiurl{10.1088/0004-637X/736/1/43}
\end{barticle}
\endbibitem

\bibitem[\protect\citeauthoryear{{Coe} et~al.}{2012}]{CO12.1}
\begin{barticle}
\bauthor{\binits{D.} \bsnm{{Coe}}},
\bauthor{\binits{K.} \bsnm{{Umetsu}}},
\bauthor{\binits{A.} \bsnm{{Zitrin}}},
\bauthor{\binits{M.} \bsnm{{Donahue}}},
\bauthor{\binits{E.} \bsnm{{Medezinski}}},
\bauthor{\binits{M.} \bsnm{{Postman}}},
\bauthor{\binits{M.} \bsnm{{Carrasco}}},
\bauthor{\binits{T.} \bsnm{{Anguita}}},
\bauthor{\binits{M.J.} \bsnm{{Geller}}},
\bauthor{\binits{K.J.} \bsnm{{Rines}}},
\bauthor{\binits{A.} \bsnm{{Diaferio}}},
\bauthor{\binits{M.J.} \bsnm{{Kurtz}}},
\bauthor{\binits{L.} \bsnm{{Bradley}}},
\bauthor{\binits{A.} \bsnm{{Koekemoer}}},
\bauthor{\binits{W.} \bsnm{{Zheng}}},
\bauthor{\binits{M.} \bsnm{{Nonino}}},
\bauthor{\binits{A.} \bsnm{{Molino}}},
\bauthor{\binits{A.} \bsnm{{Mahdavi}}},
\bauthor{\binits{D.} \bsnm{{Lemze}}},
\bauthor{\binits{L.} \bsnm{{Infante}}},
\bauthor{\binits{S.} \bsnm{{Ogaz}}},
\bauthor{\binits{P.} \bsnm{{Melchior}}},
\bauthor{\binits{O.} \bsnm{{Host}}},
\bauthor{\binits{H.} \bsnm{{Ford}}},
\bauthor{\binits{C.} \bsnm{{Grillo}}},
\bauthor{\binits{P.} \bsnm{{Rosati}}},
\bauthor{\binits{Y.} \bsnm{{Jim{\'e}nez-Teja}}},
\bauthor{\binits{J.} \bsnm{{Moustakas}}},
\bauthor{\binits{T.} \bsnm{{Broadhurst}}},
\bauthor{\binits{B.} \bsnm{{Ascaso}}},
\bauthor{\binits{O.} \bsnm{{Lahav}}},
\bauthor{\binits{M.} \bsnm{{Bartelmann}}},
\bauthor{\binits{N.} \bsnm{{Ben{\'{\i}}tez}}},
\bauthor{\binits{R.} \bsnm{{Bouwens}}},
\bauthor{\binits{O.} \bsnm{{Graur}}},
\bauthor{\binits{G.} \bsnm{{Graves}}},
\bauthor{\binits{S.} \bsnm{{Jha}}},
\bauthor{\binits{S.} \bsnm{{Jouvel}}},
\bauthor{\binits{D.} \bsnm{{Kelson}}},
\bauthor{\binits{L.} \bsnm{{Moustakas}}},
\bauthor{\binits{D.} \bsnm{{Maoz}}},
\bauthor{\binits{M.} \bsnm{{Meneghetti}}},
\bauthor{\binits{J.} \bsnm{{Merten}}},
\bauthor{\binits{A.} \bsnm{{Riess}}},
\bauthor{\binits{S.} \bsnm{{Rodney}}},
\bauthor{\binits{S.} \bsnm{{Seitz}}},
\batitle{Clash: Precise new constraints on the mass profile of the galaxy
  cluster a2261}.
\bjtitle{ApJ}
\bvolume{757},
\bfpage{22}
(\byear{2012}).
doi:\doiurl{10.1088/0004-637X/757/1/22}
\end{barticle}
\endbibitem

\bibitem[\protect\citeauthoryear{{Comerford} and {Natarajan}}{2007}]{CO07.1}
\begin{barticle}
\bauthor{\binits{J.M.} \bsnm{{Comerford}}},
\bauthor{\binits{P.} \bsnm{{Natarajan}}},
\batitle{The observed concentration-mass relation for galaxy clusters}.
\bjtitle{MNRAS}
\bvolume{379},
\bfpage{190}--\blpage{200}
(\byear{2007}).
doi:\doiurl{10.1111/j.1365-2966.2007.11934.x}
\end{barticle}
\endbibitem

\bibitem[\protect\citeauthoryear{{Davis} and {Peebles}}{1983}]{DA83.1}
\begin{barticle}
\bauthor{\binits{M.} \bsnm{{Davis}}},
\bauthor{\binits{P.J.E.} \bsnm{{Peebles}}},
\batitle{A survey of galaxy redshifts. v - the two-point position and velocity
  correlations}.
\bjtitle{ApJ}
\bvolume{267},
\bfpage{465}--\blpage{482}
(\byear{1983}).
doi:\doiurl{10.1086/160884}
\end{barticle}
\endbibitem

\bibitem[\protect\citeauthoryear{{De Grandi} and {Molendi}}{2002}]{GR02.2}
\begin{barticle}
\bauthor{\binits{S.} \bsnm{{De Grandi}}},
\bauthor{\binits{S.} \bsnm{{Molendi}}},
\batitle{Temperature profiles of nearby clusters of galaxies}.
\bjtitle{ApJ}
\bvolume{567},
\bfpage{163}--\blpage{177}
(\byear{2002}).
doi:\doiurl{10.1086/338378}
\end{barticle}
\endbibitem

\bibitem[\protect\citeauthoryear{{Dolag} et~al.}{2004}]{DO04.2}
\begin{barticle}
\bauthor{\binits{K.} \bsnm{{Dolag}}},
\bauthor{\binits{M.} \bsnm{{Bartelmann}}},
\bauthor{\binits{F.} \bsnm{{Perrotta}}},
\bauthor{\binits{C.} \bsnm{{Baccigalupi}}},
\bauthor{\binits{L.} \bsnm{{Moscardini}}},
\bauthor{\binits{M.} \bsnm{{Meneghetti}}},
\bauthor{\binits{G.} \bsnm{{Tormen}}},
\batitle{Numerical study of halo concentrations in dark-energy cosmologies}.
\bjtitle{A\&A}
\bvolume{416},
\bfpage{853}--\blpage{864}
(\byear{2004}).
doi:\doiurl{10.1051/0004-6361:20031757}
\end{barticle}
\endbibitem

\bibitem[\protect\citeauthoryear{{Dolag} et~al.}{2009}]{DO09.2}
\begin{barticle}
\bauthor{\binits{K.} \bsnm{{Dolag}}},
\bauthor{\binits{S.} \bsnm{{Borgani}}},
\bauthor{\binits{G.} \bsnm{{Murante}}},
\bauthor{\binits{V.} \bsnm{{Springel}}},
\batitle{Substructures in hydrodynamical cluster simulations}.
\bjtitle{MNRAS}
\bvolume{399},
\bfpage{497}--\blpage{514}
(\byear{2009}).
doi:\doiurl{10.1111/j.1365-2966.2009.15034.x}
\end{barticle}
\endbibitem

\bibitem[\protect\citeauthoryear{{Doroshkevich}}{1970}]{DO70.1}
\begin{barticle}
\bauthor{\binits{A.G.} \bsnm{{Doroshkevich}}},
\batitle{The space structure of perturbations and the origin of rotation of
  galaxies in the theory of fluctuation.}
\bjtitle{Astrofizika}
\bvolume{6},
\bfpage{581}--\blpage{600}
(\byear{1970})
\end{barticle}
\endbibitem

\bibitem[\protect\citeauthoryear{{Duffy} et~al.}{2008}]{DU08.1}
\begin{barticle}
\bauthor{\binits{A.R.} \bsnm{{Duffy}}},
\bauthor{\binits{J.} \bsnm{{Schaye}}},
\bauthor{\binits{S.T.} \bsnm{{Kay}}},
\bauthor{\binits{C.} \bsnm{{Dalla Vecchia}}},
\batitle{Dark matter halo concentrations in the wilkinson microwave anisotropy
  probe year 5 cosmology}.
\bjtitle{MNRAS}
\bvolume{390},
\bfpage{64}--\blpage{68}
(\byear{2008}).
doi:\doiurl{10.1111/j.1745-3933.2008.00537.x}
\end{barticle}
\endbibitem

\bibitem[\protect\citeauthoryear{{Einasto} and {Haud}}{1989}]{EI89.1}
\begin{barticle}
\bauthor{\binits{J.} \bsnm{{Einasto}}},
\bauthor{\binits{U.} \bsnm{{Haud}}},
\batitle{Galactic models with massive corona. i - method. ii - galaxy}.
\bjtitle{A\&A}
\bvolume{223},
\bfpage{89}--\blpage{106}
(\byear{1989})
\end{barticle}
\endbibitem

\bibitem[\protect\citeauthoryear{{Eke} et~al.}{2001}]{EK01.1}
\begin{barticle}
\bauthor{\binits{V.R.} \bsnm{{Eke}}},
\bauthor{\binits{J.F.} \bsnm{{Navarro}}},
\bauthor{\binits{M.} \bsnm{{Steinmetz}}},
\batitle{The power spectrum dependence of dark matter halo concentrations}.
\bjtitle{ApJ}
\bvolume{554},
\bfpage{114}--\blpage{125}
(\byear{2001}).
doi:\doiurl{10.1086/321345}
\end{barticle}
\endbibitem

\bibitem[\protect\citeauthoryear{{Ettori} et~al.}{2002}]{ET02.1}
\begin{barticle}
\bauthor{\binits{S.} \bsnm{{Ettori}}},
\bauthor{\binits{A.C.} \bsnm{{Fabian}}},
\bauthor{\binits{S.W.} \bsnm{{Allen}}},
\bauthor{\binits{R.M.} \bsnm{{Johnstone}}},
\batitle{Deep inside the core of abell 1795: the chandra view}.
\bjtitle{MNRAS}
\bvolume{331},
\bfpage{635}--\blpage{648}
(\byear{2002}).
doi:\doiurl{10.1046/j.1365-8711.2002.05212.x}
\end{barticle}
\endbibitem

\bibitem[\protect\citeauthoryear{{Ettori} et~al.}{2010}]{ET10.1}
\begin{barticle}
\bauthor{\binits{S.} \bsnm{{Ettori}}},
\bauthor{\binits{F.} \bsnm{{Gastaldello}}},
\bauthor{\binits{A.} \bsnm{{Leccardi}}},
\bauthor{\binits{S.} \bsnm{{Molendi}}},
\bauthor{\binits{M.} \bsnm{{Rossetti}}},
\bauthor{\binits{D.} \bsnm{{Buote}}},
\bauthor{\binits{M.} \bsnm{{Meneghetti}}},
\batitle{Mass profiles and c-m$_{DM}$ relation in x-ray luminous galaxy
  clusters}.
\bjtitle{A\&A}
\bvolume{524},
\bfpage{68}
(\byear{2010}).
doi:\doiurl{10.1051/0004-6361/201015271}
\end{barticle}
\endbibitem

\bibitem[\protect\citeauthoryear{{Ettori} et~al.}{2011}]{ET11.1}
\begin{barticle}
\bauthor{\binits{S.} \bsnm{{Ettori}}},
\bauthor{\binits{F.} \bsnm{{Gastaldello}}},
\bauthor{\binits{A.} \bsnm{{Leccardi}}},
\bauthor{\binits{S.} \bsnm{{Molendi}}},
\bauthor{\binits{M.} \bsnm{{Rossetti}}},
\bauthor{\binits{D.} \bsnm{{Buote}}},
\bauthor{\binits{M.} \bsnm{{Meneghetti}}},
\batitle{Mass profiles and c - m$_{DM}$ relation in x-ray luminous galaxy
  clusters (corrigendum)}.
\bjtitle{A\&A}
\bvolume{526},
\bfpage{1}
(\byear{2011}).
doi:\doiurl{10.1051/0004-6361/201015271e}
\end{barticle}
\endbibitem

\bibitem[\protect\citeauthoryear{{Fabian}}{1994}]{FA94.2}
\begin{barticle}
\bauthor{\binits{A.C.} \bsnm{{Fabian}}},
\batitle{Cooling flows in clusters of galaxies}.
\bjtitle{ARA\&A}
\bvolume{32},
\bfpage{277}--\blpage{318}
(\byear{1994}).
doi:\doiurl{10.1146/annurev.aa.32.090194.001425}
\end{barticle}
\endbibitem

\bibitem[\protect\citeauthoryear{{Fabian} et~al.}{2001}]{FA01.1}
\begin{barticle}
\bauthor{\binits{A.C.} \bsnm{{Fabian}}},
\bauthor{\binits{R.F.} \bsnm{{Mushotzky}}},
\bauthor{\binits{P.E.J.} \bsnm{{Nulsen}}},
\bauthor{\binits{J.R.} \bsnm{{Peterson}}},
\batitle{On the soft x-ray spectrum of cooling flows}.
\bjtitle{MNRAS}
\bvolume{321},
\bfpage{20}--\blpage{24}
(\byear{2001}).
doi:\doiurl{10.1046/j.1365-8711.2001.04285.x}
\end{barticle}
\endbibitem

\bibitem[\protect\citeauthoryear{{Fabian} et~al.}{2011}]{FA11.1}
\begin{barticle}
\bauthor{\binits{A.C.} \bsnm{{Fabian}}},
\bauthor{\binits{J.S.} \bsnm{{Sanders}}},
\bauthor{\binits{S.W.} \bsnm{{Allen}}},
\bauthor{\binits{R.E.A.} \bsnm{{Canning}}},
\bauthor{\binits{E.} \bsnm{{Churazov}}},
\bauthor{\binits{C.S.} \bsnm{{Crawford}}},
\bauthor{\binits{W.} \bsnm{{Forman}}},
\bauthor{\binits{J.} \bsnm{{Gabany}}},
\bauthor{\binits{J.} \bsnm{{Hlavacek-Larrondo}}},
\bauthor{\binits{R.M.} \bsnm{{Johnstone}}},
\bauthor{\binits{H.R.} \bsnm{{Russell}}},
\bauthor{\binits{C.S.} \bsnm{{Reynolds}}},
\bauthor{\binits{P.} \bsnm{{Salom{\'e}}}},
\bauthor{\binits{G.B.} \bsnm{{Taylor}}},
\bauthor{\binits{A.J.} \bsnm{{Young}}},
\batitle{A wide chandra view of the core of the perseus cluster}.
\bjtitle{MNRAS}
\bvolume{418},
\bfpage{2154}--\blpage{2164}
(\byear{2011}).
doi:\doiurl{10.1111/j.1365-2966.2011.19402.x}
\end{barticle}
\endbibitem

\bibitem[\protect\citeauthoryear{{Gao} et~al.}{2004}]{GA04.2}
\begin{barticle}
\bauthor{\binits{L.} \bsnm{{Gao}}},
\bauthor{\binits{S.D.M.} \bsnm{{White}}},
\bauthor{\binits{A.} \bsnm{{Jenkins}}},
\bauthor{\binits{F.} \bsnm{{Stoehr}}},
\bauthor{\binits{V.} \bsnm{{Springel}}},
\batitle{The subhalo populations of {$\Lambda$}cdm dark haloes}.
\bjtitle{MNRAS}
\bvolume{355},
\bfpage{819}--\blpage{834}
(\byear{2004}).
doi:\doiurl{10.1111/j.1365-2966.2004.08360.x}
\end{barticle}
\endbibitem

\bibitem[\protect\citeauthoryear{{Gao} et~al.}{2008}]{GA08.1}
\begin{barticle}
\bauthor{\binits{L.} \bsnm{{Gao}}},
\bauthor{\binits{J.F.} \bsnm{{Navarro}}},
\bauthor{\binits{S.} \bsnm{{Cole}}},
\bauthor{\binits{C.S.} \bsnm{{Frenk}}},
\bauthor{\binits{S.D.M.} \bsnm{{White}}},
\bauthor{\binits{V.} \bsnm{{Springel}}},
\bauthor{\binits{A.} \bsnm{{Jenkins}}},
\bauthor{\binits{A.F.} \bsnm{{Neto}}},
\batitle{The redshift dependence of the structure of massive {$\Lambda$} cold
  dark matter haloes}.
\bjtitle{MNRAS}
\bvolume{387},
\bfpage{536}--\blpage{544}
(\byear{2008}).
doi:\doiurl{10.1111/j.1365-2966.2008.13277.x}
\end{barticle}
\endbibitem

\bibitem[\protect\citeauthoryear{{Gao} et~al.}{2011}]{GA11.1}
\begin{barticle}
\bauthor{\binits{L.} \bsnm{{Gao}}},
\bauthor{\binits{C.S.} \bsnm{{Frenk}}},
\bauthor{\binits{M.} \bsnm{{Boylan-Kolchin}}},
\bauthor{\binits{A.} \bsnm{{Jenkins}}},
\bauthor{\binits{V.} \bsnm{{Springel}}},
\bauthor{\binits{S.D.M.} \bsnm{{White}}},
\batitle{The statistics of the subhalo abundance of dark matter haloes}.
\bjtitle{MNRAS}
\bvolume{410},
\bfpage{2309}--\blpage{2314}
(\byear{2011}).
doi:\doiurl{10.1111/j.1365-2966.2010.17601.x}
\end{barticle}
\endbibitem

\bibitem[\protect\citeauthoryear{{Gao} et~al.}{2012}]{GA12.1}
\begin{barticle}
\bauthor{\binits{L.} \bsnm{{Gao}}},
\bauthor{\binits{C.S.} \bsnm{{Frenk}}},
\bauthor{\binits{A.} \bsnm{{Jenkins}}},
\bauthor{\binits{V.} \bsnm{{Springel}}},
\bauthor{\binits{S.D.M.} \bsnm{{White}}},
\batitle{Where will supersymmetric dark matter first be seen?}
\bjtitle{MNRAS}
\bvolume{419},
\bfpage{1721}--\blpage{1726}
(\byear{2012}).
doi:\doiurl{10.1111/j.1365-2966.2011.19836.x}
\end{barticle}
\endbibitem

\bibitem[\protect\citeauthoryear{{Gastaldello} et~al.}{2007}]{GA07.3}
\begin{barticle}
\bauthor{\binits{F.} \bsnm{{Gastaldello}}},
\bauthor{\binits{D.A.} \bsnm{{Buote}}},
\bauthor{\binits{P.J.} \bsnm{{Humphrey}}},
\bauthor{\binits{L.} \bsnm{{Zappacosta}}},
\bauthor{\binits{J.S.} \bsnm{{Bullock}}},
\bauthor{\binits{F.} \bsnm{{Brighenti}}},
\bauthor{\binits{W.G.} \bsnm{{Mathews}}},
\batitle{Probing the dark matter and gas fraction in relaxed galaxy groups with
  x-ray observations from chandra and xmm-newton}.
\bjtitle{ApJ}
\bvolume{669},
\bfpage{158}--\blpage{183}
(\byear{2007}).
doi:\doiurl{10.1086/521519}
\end{barticle}
\endbibitem

\bibitem[\protect\citeauthoryear{{Giocoli} et~al.}{2010a}]{GI10.1}
\begin{barticle}
\bauthor{\binits{C.} \bsnm{{Giocoli}}},
\bauthor{\binits{M.} \bsnm{{Bartelmann}}},
\bauthor{\binits{R.K.} \bsnm{{Sheth}}},
\bauthor{\binits{M.} \bsnm{{Cacciato}}},
\batitle{Halo model description of the non-linear dark matter power spectrum at
  $k \gg 1\,\mathrm{Mpc}^{-1}$}.
\bjtitle{MNRAS}
\bvolume{408},
\bfpage{300}--\blpage{313}
(\byear{2010}a).
doi:\doiurl{10.1111/j.1365-2966.2010.17108.x}
\end{barticle}
\endbibitem

\bibitem[\protect\citeauthoryear{{Giocoli} et~al.}{2010b}]{GI10.2}
\begin{barticle}
\bauthor{\binits{C.} \bsnm{{Giocoli}}},
\bauthor{\binits{G.} \bsnm{{Tormen}}},
\bauthor{\binits{R.K.} \bsnm{{Sheth}}},
\bauthor{\binits{F.C.} \bsnm{{van den Bosch}}},
\batitle{The substructure hierarchy in dark matter haloes}.
\bjtitle{MNRAS}
\bvolume{404},
\bfpage{502}--\blpage{517}
(\byear{2010}b).
doi:\doiurl{10.1111/j.1365-2966.2010.16311.x}
\end{barticle}
\endbibitem

\bibitem[\protect\citeauthoryear{{Gnedin} et~al.}{2004}]{GN04.1}
\begin{barticle}
\bauthor{\binits{O.Y.} \bsnm{{Gnedin}}},
\bauthor{\binits{A.V.} \bsnm{{Kravtsov}}},
\bauthor{\binits{A.A.} \bsnm{{Klypin}}},
\bauthor{\binits{D.} \bsnm{{Nagai}}},
\batitle{Response of dark matter halos to condensation of baryons: Cosmological
  simulations and improved adiabatic contraction model}.
\bjtitle{ApJ}
\bvolume{616},
\bfpage{16}--\blpage{26}
(\byear{2004}).
doi:\doiurl{10.1086/424914}
\end{barticle}
\endbibitem

\bibitem[\protect\citeauthoryear{{Goldberg} and {Bacon}}{2005}]{GO05.1}
\begin{barticle}
\bauthor{\binits{D.M.} \bsnm{{Goldberg}}},
\bauthor{\binits{D.J.} \bsnm{{Bacon}}},
\batitle{Galaxy-galaxy flexion: Weak lensing to second order}.
\bjtitle{ApJ}
\bvolume{619},
\bfpage{741}--\blpage{748}
(\byear{2005}).
doi:\doiurl{10.1086/426782}
\end{barticle}
\endbibitem

\bibitem[\protect\citeauthoryear{{Goldberg} and {Leonard}}{2007}]{GO07.1}
\begin{barticle}
\bauthor{\binits{D.M.} \bsnm{{Goldberg}}},
\bauthor{\binits{A.} \bsnm{{Leonard}}},
\batitle{Measuring flexion}.
\bjtitle{ApJ}
\bvolume{660},
\bfpage{1003}--\blpage{1015}
(\byear{2007}).
doi:\doiurl{10.1086/513137}
\end{barticle}
\endbibitem

\bibitem[\protect\citeauthoryear{{Halkola} et~al.}{2006}]{HA06.1}
\begin{barticle}
\bauthor{\binits{A.} \bsnm{{Halkola}}},
\bauthor{\binits{S.} \bsnm{{Seitz}}},
\bauthor{\binits{M.} \bsnm{{Pannella}}},
\batitle{Parametric strong gravitational lensing analysis of abell 1689}.
\bjtitle{MNRAS}
\bvolume{372},
\bfpage{1425}--\blpage{1462}
(\byear{2006}).
doi:\doiurl{10.1111/j.1365-2966.2006.10948.x}
\end{barticle}
\endbibitem

\bibitem[\protect\citeauthoryear{{Hennawi} et~al.}{2007}]{HE07.1}
\begin{barticle}
\bauthor{\binits{J.F.} \bsnm{{Hennawi}}},
\bauthor{\binits{N.} \bsnm{{Dalal}}},
\bauthor{\binits{P.} \bsnm{{Bode}}},
\bauthor{\binits{J.P.} \bsnm{{Ostriker}}},
\batitle{Characterizing the cluster lens population}.
\bjtitle{ApJ}
\bvolume{654},
\bfpage{714}--\blpage{730}
(\byear{2007}).
doi:\doiurl{10.1086/497362}
\end{barticle}
\endbibitem

\bibitem[\protect\citeauthoryear{{Host} and {Hansen}}{2011}]{HO11.2}
\begin{barticle}
\bauthor{\binits{O.} \bsnm{{Host}}},
\bauthor{\binits{S.H.} \bsnm{{Hansen}}},
\batitle{A detailed statistical analysis of the mass profiles of galaxy
  clusters}.
\bjtitle{ApJ}
\bvolume{736},
\bfpage{52}
(\byear{2011}).
doi:\doiurl{10.1088/0004-637X/736/1/52}
\end{barticle}
\endbibitem

\bibitem[\protect\citeauthoryear{{Huang} et~al.}{2012}]{HU12.1}
\begin{barticle}
\bauthor{\binits{X.} \bsnm{{Huang}}},
\bauthor{\binits{G.} \bsnm{{Vertongen}}},
\bauthor{\binits{C.} \bsnm{{Weniger}}},
\batitle{Probing dark matter decay and annihilation with fermi lat observations
  of nearby galaxy clusters}.
\bjtitle{JCAP}
\bvolume{1},
\bfpage{42}
(\byear{2012}).
doi:\doiurl{10.1088/1475-7516/2012/01/042}
\end{barticle}
\endbibitem

\bibitem[\protect\citeauthoryear{{Jing} and {Suto}}{2000}]{JI00.1}
\begin{barticle}
\bauthor{\binits{Y.P.} \bsnm{{Jing}}},
\bauthor{\binits{Y.} \bsnm{{Suto}}},
\batitle{The density profiles of the dark matter halo are not universal}.
\bjtitle{ApJl}
\bvolume{529},
\bfpage{69}--\blpage{72}
(\byear{2000}).
doi:\doiurl{10.1086/312463}
\end{barticle}
\endbibitem

\bibitem[\protect\citeauthoryear{{Komatsu} et~al.}{2011}]{KO11.1}
\begin{barticle}
\bauthor{\binits{E.} \bsnm{{Komatsu}}},
\bauthor{\binits{K.M.} \bsnm{{Smith}}},
\bauthor{\binits{J.} \bsnm{{Dunkley}}},
\bauthor{\binits{C.L.} \bsnm{{Bennett}}},
\bauthor{\binits{B.} \bsnm{{Gold}}},
\bauthor{\binits{G.} \bsnm{{Hinshaw}}},
\bauthor{\binits{N.} \bsnm{{Jarosik}}},
\bauthor{\binits{D.} \bsnm{{Larson}}},
\bauthor{\binits{M.R.} \bsnm{{Nolta}}},
\bauthor{\binits{L.} \bsnm{{Page}}},
\bauthor{\binits{D.N.} \bsnm{{Spergel}}},
\bauthor{\binits{M.} \bsnm{{Halpern}}},
\bauthor{\binits{R.S.} \bsnm{{Hill}}},
\bauthor{\binits{A.} \bsnm{{Kogut}}},
\bauthor{\binits{M.} \bsnm{{Limon}}},
\bauthor{\binits{S.S.} \bsnm{{Meyer}}},
\bauthor{\binits{N.} \bsnm{{Odegard}}},
\bauthor{\binits{G.S.} \bsnm{{Tucker}}},
\bauthor{\binits{J.L.} \bsnm{{Weiland}}},
\bauthor{\binits{E.} \bsnm{{Wollack}}},
\bauthor{\binits{E.L.} \bsnm{{Wright}}},
\batitle{Seven-year wilkinson microwave anisotropy probe (wmap) observations:
  Cosmological interpretation}.
\bjtitle{ApJs}
\bvolume{192},
\bfpage{18}
(\byear{2011}).
doi:\doiurl{10.1088/0067-0049/192/2/18}
\end{barticle}
\endbibitem

\bibitem[\protect\citeauthoryear{{Lemze} et~al.}{2009}]{LE09.1}
\begin{barticle}
\bauthor{\binits{D.} \bsnm{{Lemze}}},
\bauthor{\binits{T.} \bsnm{{Broadhurst}}},
\bauthor{\binits{Y.} \bsnm{{Rephaeli}}},
\bauthor{\binits{R.} \bsnm{{Barkana}}},
\bauthor{\binits{K.} \bsnm{{Umetsu}}},
\batitle{Dynamical study of a1689 from wide-field vlt/vimos spectroscopy: Mass
  profile, concentration parameter, and velocity anisotropy}.
\bjtitle{ApJ}
\bvolume{701},
\bfpage{1336}--\blpage{1346}
(\byear{2009}).
doi:\doiurl{10.1088/0004-637X/701/2/1336}
\end{barticle}
\endbibitem

\bibitem[\protect\citeauthoryear{{Leonard} et~al.}{2011}]{LE11.2}
\begin{barticle}
\bauthor{\binits{A.} \bsnm{{Leonard}}},
\bauthor{\binits{L.J.} \bsnm{{King}}},
\bauthor{\binits{D.M.} \bsnm{{Goldberg}}},
\batitle{New constraints on the complex mass substructure in abell 1689 from
  gravitational flexion}.
\bjtitle{MNRAS}
\bvolume{413},
\bfpage{789}--\blpage{804}
(\byear{2011}).
doi:\doiurl{10.1111/j.1365-2966.2010.18171.x}
\end{barticle}
\endbibitem

\bibitem[\protect\citeauthoryear{{Leonard} et~al.}{2007}]{LE07.1}
\begin{barticle}
\bauthor{\binits{A.} \bsnm{{Leonard}}},
\bauthor{\binits{D.M.} \bsnm{{Goldberg}}},
\bauthor{\binits{J.L.} \bsnm{{Haaga}}},
\bauthor{\binits{R.} \bsnm{{Massey}}},
\batitle{Gravitational shear, flexion, and strong lensing in abell 1689}.
\bjtitle{ApJ}
\bvolume{666},
\bfpage{51}--\blpage{63}
(\byear{2007}).
doi:\doiurl{10.1086/520109}
\end{barticle}
\endbibitem

\bibitem[\protect\citeauthoryear{{Lerchster} et~al.}{2011}]{LE11.1}
\begin{barticle}
\bauthor{\binits{M.} \bsnm{{Lerchster}}},
\bauthor{\binits{S.} \bsnm{{Seitz}}},
\bauthor{\binits{F.} \bsnm{{Brimioulle}}},
\bauthor{\binits{R.} \bsnm{{Fassbender}}},
\bauthor{\binits{M.} \bsnm{{Rovilos}}},
\bauthor{\binits{H.} \bsnm{{B{\"o}hringer}}},
\bauthor{\binits{D.} \bsnm{{Pierini}}},
\bauthor{\binits{M.} \bsnm{{Kilbinger}}},
\bauthor{\binits{A.} \bsnm{{Finoguenov}}},
\bauthor{\binits{H.} \bsnm{{Quintana}}},
\bauthor{\binits{R.} \bsnm{{Bender}}},
\batitle{The massive galaxy cluster xmmu j1230.3+1339 at $z \sim 1$:
  colour-magnitude relation, butcher-oemler effect, x-ray and weak lensing mass
  estimates}.
\bjtitle{MNRAS}
\bvolume{411},
\bfpage{2667}--\blpage{2694}
(\byear{2011}).
doi:\doiurl{10.1111/j.1365-2966.2010.17874.x}
\end{barticle}
\endbibitem

\bibitem[\protect\citeauthoryear{{Lewis} et~al.}{2003}]{LE03.1}
\begin{barticle}
\bauthor{\binits{A.D.} \bsnm{{Lewis}}},
\bauthor{\binits{D.A.} \bsnm{{Buote}}},
\bauthor{\binits{J.T.} \bsnm{{Stocke}}},
\batitle{Chandra observations of a2029: The dark matter profile down to below
  0.01r$_{vir}$ in an unusually relaxed cluster}.
\bjtitle{ApJ}
\bvolume{586},
\bfpage{135}--\blpage{142}
(\byear{2003}).
doi:\doiurl{10.1086/367556}
\end{barticle}
\endbibitem

\bibitem[\protect\citeauthoryear{{Limousin} et~al.}{2008}]{LI08.1}
\begin{barticle}
\bauthor{\binits{M.} \bsnm{{Limousin}}},
\bauthor{\binits{J.} \bsnm{{Richard}}},
\bauthor{\binits{J.} \bsnm{{Kneib}}},
\bauthor{\binits{H.} \bsnm{{Brink}}},
\bauthor{\binits{R.} \bsnm{{Pell{\'o}}}},
\bauthor{\binits{E.} \bsnm{{Jullo}}},
\bauthor{\binits{H.} \bsnm{{Tu}}},
\bauthor{\binits{J.} \bsnm{{Sommer-Larsen}}},
\bauthor{\binits{E.} \bsnm{{Egami}}},
\bauthor{\binits{M.J.} \bsnm{{Micha{\l}owski}}},
\bauthor{\binits{R.} \bsnm{{Cabanac}}},
\bauthor{\binits{D.P.} \bsnm{{Stark}}},
\batitle{Strong lensing in abell 1703: constraints on the slope of the inner
  dark matter distribution}.
\bjtitle{A\&A}
\bvolume{489},
\bfpage{23}--\blpage{35}
(\byear{2008}).
doi:\doiurl{10.1051/0004-6361:200809646}
\end{barticle}
\endbibitem

\bibitem[\protect\citeauthoryear{{Macci{\`o}} et~al.}{2008}]{MA08.2}
\begin{barticle}
\bauthor{\binits{A.V.} \bsnm{{Macci{\`o}}}},
\bauthor{\binits{A.A.} \bsnm{{Dutton}}},
\bauthor{\binits{F.C.} \bsnm{{van den Bosch}}},
\batitle{Concentration, spin and shape of dark matter haloes as a function of
  the cosmological model: Wmap1, wmap3 and wmap5 results}.
\bjtitle{MNRAS}
\bvolume{391},
\bfpage{1940}--\blpage{1954}
(\byear{2008}).
doi:\doiurl{10.1111/j.1365-2966.2008.14029.x}
\end{barticle}
\endbibitem

\bibitem[\protect\citeauthoryear{{Macci{\`o}} et~al.}{2007}]{MA07.5}
\begin{barticle}
\bauthor{\binits{A.V.} \bsnm{{Macci{\`o}}}},
\bauthor{\binits{A.A.} \bsnm{{Dutton}}},
\bauthor{\binits{F.C.} \bsnm{{van den Bosch}}},
\bauthor{\binits{B.} \bsnm{{Moore}}},
\bauthor{\binits{D.} \bsnm{{Potter}}},
\bauthor{\binits{J.} \bsnm{{Stadel}}},
\batitle{Concentration, spin and shape of dark matter haloes: scatter and the
  dependence on mass and environment}.
\bjtitle{MNRAS}
\bvolume{378},
\bfpage{55}--\blpage{71}
(\byear{2007}).
doi:\doiurl{10.1111/j.1365-2966.2007.11720.x}
\end{barticle}
\endbibitem

\bibitem[\protect\citeauthoryear{{Mandelbaum} et~al.}{2008}]{MA08.1}
\begin{barticle}
\bauthor{\binits{R.} \bsnm{{Mandelbaum}}},
\bauthor{\binits{U.} \bsnm{{Seljak}}},
\bauthor{\binits{C.M.} \bsnm{{Hirata}}},
\batitle{A halo mass-concentration relation from weak lensing}.
\bjtitle{Journal of Cosmology and Astro-Particle Physics}
\bvolume{8},
\bfpage{6}
(\byear{2008}).
doi:\doiurl{10.1088/1475-7516/2008/08/006}
\end{barticle}
\endbibitem

\bibitem[\protect\citeauthoryear{{Mead} et~al.}{2010}]{ME10.3}
\begin{barticle}
\bauthor{\binits{J.M.G.} \bsnm{{Mead}}},
\bauthor{\binits{L.J.} \bsnm{{King}}},
\bauthor{\binits{D.} \bsnm{{Sijacki}}},
\bauthor{\binits{A.} \bsnm{{Leonard}}},
\bauthor{\binits{E.} \bsnm{{Puchwein}}},
\bauthor{\binits{I.G.} \bsnm{{McCarthy}}},
\batitle{The impact of agn feedback and baryonic cooling on galaxy clusters as
  gravitational lenses}.
\bjtitle{MNRAS}
\bvolume{406},
\bfpage{434}--\blpage{444}
(\byear{2010}).
doi:\doiurl{10.1111/j.1365-2966.2010.16674.x}
\end{barticle}
\endbibitem

\bibitem[\protect\citeauthoryear{{Meneghetti} et~al.}{2010a}]{ME10.1}
\begin{barticle}
\bauthor{\binits{M.} \bsnm{{Meneghetti}}},
\bauthor{\binits{C.} \bsnm{{Fedeli}}},
\bauthor{\binits{F.} \bsnm{{Pace}}},
\bauthor{\binits{S.} \bsnm{{Gottl{\"o}ber}}},
\bauthor{\binits{G.} \bsnm{{Yepes}}},
\batitle{Strong lensing in the marenostrum universe. i. biases in the cluster
  lens population}.
\bjtitle{A\&A}
\bvolume{519},
\bfpage{90}
(\byear{2010}a).
doi:\doiurl{10.1051/0004-6361/201014098}
\end{barticle}
\endbibitem

\bibitem[\protect\citeauthoryear{{Meneghetti} et~al.}{2010b}]{ME10.2}
\begin{barticle}
\bauthor{\binits{M.} \bsnm{{Meneghetti}}},
\bauthor{\binits{E.} \bsnm{{Rasia}}},
\bauthor{\binits{J.} \bsnm{{Merten}}},
\bauthor{\binits{F.} \bsnm{{Bellagamba}}},
\bauthor{\binits{S.} \bsnm{{Ettori}}},
\bauthor{\binits{P.} \bsnm{{Mazzotta}}},
\bauthor{\binits{K.} \bsnm{{Dolag}}},
\bauthor{\binits{S.} \bsnm{{Marri}}},
\batitle{Weighing simulated galaxy clusters using lensing and x-ray}.
\bjtitle{A\&A}
\bvolume{514},
\bfpage{93}
(\byear{2010}b).
doi:\doiurl{10.1051/0004-6361/200913222}
\end{barticle}
\endbibitem

\bibitem[\protect\citeauthoryear{{Merritt} et~al.}{2006}]{ME06.1}
\begin{barticle}
\bauthor{\binits{D.} \bsnm{{Merritt}}},
\bauthor{\binits{A.W.} \bsnm{{Graham}}},
\bauthor{\binits{B.} \bsnm{{Moore}}},
\bauthor{\binits{J.} \bsnm{{Diemand}}},
\bauthor{\binits{B.} \bsnm{{Terzi{\'c}}}},
\batitle{Empirical models for dark matter halos. i. nonparametric construction
  of density profiles and comparison with parametric models}.
\bjtitle{AJ}
\bvolume{132},
\bfpage{2685}--\blpage{2700}
(\byear{2006}).
doi:\doiurl{10.1086/508988}
\end{barticle}
\endbibitem

\bibitem[\protect\citeauthoryear{{Merten} et~al.}{2009}]{ME09.2}
\begin{barticle}
\bauthor{\binits{J.} \bsnm{{Merten}}},
\bauthor{\binits{M.} \bsnm{{Cacciato}}},
\bauthor{\binits{M.} \bsnm{{Meneghetti}}},
\bauthor{\binits{C.} \bsnm{{Mignone}}},
\bauthor{\binits{M.} \bsnm{{Bartelmann}}},
\batitle{Combining weak and strong cluster lensing: applications to simulations
  and ms 2137}.
\bjtitle{A\&A}
\bvolume{500},
\bfpage{681}--\blpage{691}
(\byear{2009}).
doi:\doiurl{10.1051/0004-6361/200810372}
\end{barticle}
\endbibitem

\bibitem[\protect\citeauthoryear{{Merten} et~al.}{2011}]{ME11.1}
\begin{barticle}
\bauthor{\binits{J.} \bsnm{{Merten}}},
\bauthor{\binits{D.} \bsnm{{Coe}}},
\bauthor{\binits{R.} \bsnm{{Dupke}}},
\bauthor{\binits{R.} \bsnm{{Massey}}},
\bauthor{\binits{A.} \bsnm{{Zitrin}}},
\bauthor{\binits{E.S.} \bsnm{{Cypriano}}},
\bauthor{\binits{N.} \bsnm{{Okabe}}},
\bauthor{\binits{B.} \bsnm{{Frye}}},
\bauthor{\binits{F.G.} \bsnm{{Braglia}}},
\bauthor{\binits{Y.} \bsnm{{Jim{\'e}nez-Teja}}},
\bauthor{\binits{N.} \bsnm{{Ben{\'{\i}}tez}}},
\bauthor{\binits{T.} \bsnm{{Broadhurst}}},
\bauthor{\binits{J.} \bsnm{{Rhodes}}},
\bauthor{\binits{M.} \bsnm{{Meneghetti}}},
\bauthor{\binits{L.A.} \bsnm{{Moustakas}}},
\bauthor{\binits{L.} \bsnm{{Sodr{\'e}}} \bsuffix{Jr.}},
\bauthor{\binits{J.} \bsnm{{Krick}}},
\bauthor{\binits{J.N.} \bsnm{{Bregman}}},
\batitle{Creation of cosmic structure in the complex galaxy cluster merger
  abell 2744}.
\bjtitle{MNRAS}
\bvolume{417},
\bfpage{333}--\blpage{347}
(\byear{2011}).
doi:\doiurl{10.1111/j.1365-2966.2011.19266.x}
\end{barticle}
\endbibitem

\bibitem[\protect\citeauthoryear{{Moore} et~al.}{1998}]{MO98.3}
\begin{barticle}
\bauthor{\binits{B.} \bsnm{{Moore}}},
\bauthor{\binits{F.} \bsnm{{Governato}}},
\bauthor{\binits{T.} \bsnm{{Quinn}}},
\bauthor{\binits{J.} \bsnm{{Stadel}}},
\bauthor{\binits{G.} \bsnm{{Lake}}},
\batitle{Resolving the structure of cold dark matter halos}.
\bjtitle{ApJl}
\bvolume{499},
\bfpage{5}
(\byear{1998}).
doi:\doiurl{10.1086/311333}
\end{barticle}
\endbibitem

\bibitem[\protect\citeauthoryear{{Moore} et~al.}{1999}]{MO99.2}
\begin{barticle}
\bauthor{\binits{B.} \bsnm{{Moore}}},
\bauthor{\binits{T.} \bsnm{{Quinn}}},
\bauthor{\binits{F.} \bsnm{{Governato}}},
\bauthor{\binits{J.} \bsnm{{Stadel}}},
\bauthor{\binits{G.} \bsnm{{Lake}}},
\batitle{Cold collapse and the core catastrophe}.
\bjtitle{MNRAS}
\bvolume{310},
\bfpage{1147}--\blpage{1152}
(\byear{1999}).
doi:\doiurl{10.1046/j.1365-8711.1999.03039.x}
\end{barticle}
\endbibitem

\bibitem[\protect\citeauthoryear{{Navarro} et~al.}{1996}]{NA96.1}
\begin{barticle}
\bauthor{\binits{J.F.} \bsnm{{Navarro}}},
\bauthor{\binits{C.S.} \bsnm{{Frenk}}},
\bauthor{\binits{S.D.M.} \bsnm{{White}}},
\batitle{The structure of cold dark matter halos}.
\bjtitle{ApJ}
\bvolume{462},
\bfpage{563}
(\byear{1996})
\end{barticle}
\endbibitem

\bibitem[\protect\citeauthoryear{{Navarro} et~al.}{1997}]{NA97.1}
\begin{barticle}
\bauthor{\binits{J.F.} \bsnm{{Navarro}}},
\bauthor{\binits{C.S.} \bsnm{{Frenk}}},
\bauthor{\binits{S.D.M.} \bsnm{{White}}},
\batitle{A universal density profile from hierarchical clustering}.
\bjtitle{ApJ}
\bvolume{490},
\bfpage{493}
(\byear{1997})
\end{barticle}
\endbibitem

\bibitem[\protect\citeauthoryear{{Navarro} et~al.}{2004}]{NA04.1}
\begin{barticle}
\bauthor{\binits{J.F.} \bsnm{{Navarro}}},
\bauthor{\binits{E.} \bsnm{{Hayashi}}},
\bauthor{\binits{C.} \bsnm{{Power}}},
\bauthor{\binits{A.R.} \bsnm{{Jenkins}}},
\bauthor{\binits{C.S.} \bsnm{{Frenk}}},
\bauthor{\binits{S.D.M.} \bsnm{{White}}},
\bauthor{\binits{V.} \bsnm{{Springel}}},
\bauthor{\binits{J.} \bsnm{{Stadel}}},
\bauthor{\binits{T.R.} \bsnm{{Quinn}}},
\batitle{The inner structure of {$\Lambda$}cdm haloes - iii. universality and
  asymptotic slopes}.
\bjtitle{MNRAS}
\bvolume{349},
\bfpage{1039}--\blpage{1051}
(\byear{2004}).
doi:\doiurl{10.1111/j.1365-2966.2004.07586.x}
\end{barticle}
\endbibitem

\bibitem[\protect\citeauthoryear{{Neto} et~al.}{2007}]{NE07.1}
\begin{barticle}
\bauthor{\binits{A.F.} \bsnm{{Neto}}},
\bauthor{\binits{L.} \bsnm{{Gao}}},
\bauthor{\binits{P.} \bsnm{{Bett}}},
\bauthor{\binits{S.} \bsnm{{Cole}}},
\bauthor{\binits{J.F.} \bsnm{{Navarro}}},
\bauthor{\binits{C.S.} \bsnm{{Frenk}}},
\bauthor{\binits{S.D.M.} \bsnm{{White}}},
\bauthor{\binits{V.} \bsnm{{Springel}}},
\bauthor{\binits{A.} \bsnm{{Jenkins}}},
\batitle{The statistics of {$\Lambda$} cdm halo concentrations}.
\bjtitle{MNRAS}
\bvolume{381},
\bfpage{1450}--\blpage{1462}
(\byear{2007}).
doi:\doiurl{10.1111/j.1365-2966.2007.12381.x}
\end{barticle}
\endbibitem

\bibitem[\protect\citeauthoryear{{Newman} et~al.}{2013}]{NE13.1}
\begin{barticle}
\bauthor{\binits{A.B.} \bsnm{{Newman}}},
\bauthor{\binits{T.} \bsnm{{Treu}}},
\bauthor{\binits{R.S.} \bsnm{{Ellis}}},
\bauthor{\binits{D.J.} \bsnm{{Sand}}},
\batitle{The density profiles of massive, relaxed galaxy clusters. ii.
  separating luminous and dark matter in cluster cores}.
\bjtitle{ApJ}
\bvolume{765},
\bfpage{25}
(\byear{2013}).
doi:\doiurl{10.1088/0004-637X/765/1/25}
\end{barticle}
\endbibitem

\bibitem[\protect\citeauthoryear{{Okabe} et~al.}{2010}]{OK10.1}
\begin{barticle}
\bauthor{\binits{N.} \bsnm{{Okabe}}},
\bauthor{\binits{M.} \bsnm{{Takada}}},
\bauthor{\binits{K.} \bsnm{{Umetsu}}},
\bauthor{\binits{T.} \bsnm{{Futamase}}},
\bauthor{\binits{G.P.} \bsnm{{Smith}}},
\batitle{Locuss: Subaru weak lensing study of 30 galaxy clusters}.
\bjtitle{PASJ}
\bvolume{62},
\bfpage{811}
(\byear{2010})
\end{barticle}
\endbibitem

\bibitem[\protect\citeauthoryear{{Okura} et~al.}{2007}]{OK07.1}
\begin{barticle}
\bauthor{\binits{Y.} \bsnm{{Okura}}},
\bauthor{\binits{K.} \bsnm{{Umetsu}}},
\bauthor{\binits{T.} \bsnm{{Futamase}}},
\batitle{A new measure for weak-lensing flexion}.
\bjtitle{ApJ}
\bvolume{660},
\bfpage{995}--\blpage{1002}
(\byear{2007}).
doi:\doiurl{10.1086/513135}
\end{barticle}
\endbibitem

\bibitem[\protect\citeauthoryear{{Okura} et~al.}{2008}]{OK08.1}
\begin{barticle}
\bauthor{\binits{Y.} \bsnm{{Okura}}},
\bauthor{\binits{K.} \bsnm{{Umetsu}}},
\bauthor{\binits{T.} \bsnm{{Futamase}}},
\batitle{A method for weak-lensing flexion analysis by the holics moment
  approach}.
\bjtitle{ApJ}
\bvolume{680},
\bfpage{1}--\blpage{16}
(\byear{2008}).
doi:\doiurl{10.1086/587676}
\end{barticle}
\endbibitem

\bibitem[\protect\citeauthoryear{{Peterson} et~al.}{2003}]{PE03.1}
\begin{barticle}
\bauthor{\binits{J.R.} \bsnm{{Peterson}}},
\bauthor{\binits{S.M.} \bsnm{{Kahn}}},
\bauthor{\binits{F.B.S.} \bsnm{{Paerels}}},
\bauthor{\binits{J.S.} \bsnm{{Kaastra}}},
\bauthor{\binits{T.} \bsnm{{Tamura}}},
\bauthor{\binits{J.A.M.} \bsnm{{Bleeker}}},
\bauthor{\binits{C.} \bsnm{{Ferrigno}}},
\bauthor{\binits{J.G.} \bsnm{{Jernigan}}},
\batitle{High-resolution x-ray spectroscopic constraints on cooling-flow models
  for clusters of galaxies}.
\bjtitle{ApJ}
\bvolume{590},
\bfpage{207}--\blpage{224}
(\byear{2003}).
doi:\doiurl{10.1086/374830}
\end{barticle}
\endbibitem

\bibitem[\protect\citeauthoryear{{Pinzke} et~al.}{2011}]{PI11.1}
\begin{barticle}
\bauthor{\binits{A.} \bsnm{{Pinzke}}},
\bauthor{\binits{C.} \bsnm{{Pfrommer}}},
\bauthor{\binits{L.} \bsnm{{Bergstr{\"o}m}}},
\batitle{Prospects of detecting gamma-ray emission from galaxy clusters: Cosmic
  rays and dark matter annihilations}.
\bjtitle{PRD}
\bvolume{84}(\bissue{12}),
\bfpage{123509}
(\byear{2011}).
doi:\doiurl{10.1103/PhysRevD.84.123509}
\end{barticle}
\endbibitem

\bibitem[\protect\citeauthoryear{{Pires} and {Amara}}{2010}]{PI10.1}
\begin{barticle}
\bauthor{\binits{S.} \bsnm{{Pires}}},
\bauthor{\binits{A.} \bsnm{{Amara}}},
\batitle{Weak lensing mass reconstruction: Flexion versus shear}.
\bjtitle{ApJ}
\bvolume{723},
\bfpage{1507}--\blpage{1511}
(\byear{2010}).
doi:\doiurl{10.1088/0004-637X/723/2/1507}
\end{barticle}
\endbibitem

\bibitem[\protect\citeauthoryear{{Pointecouteau} et~al.}{2005}]{PO05.1}
\begin{barticle}
\bauthor{\binits{E.} \bsnm{{Pointecouteau}}},
\bauthor{\binits{M.} \bsnm{{Arnaud}}},
\bauthor{\binits{G.W.} \bsnm{{Pratt}}},
\batitle{The structural and scaling properties of nearby galaxy clusters. i.
  the universal mass profile}.
\bjtitle{A\&A}
\bvolume{435},
\bfpage{1}--\blpage{7}
(\byear{2005}).
doi:\doiurl{10.1051/0004-6361:20042569}
\end{barticle}
\endbibitem

\bibitem[\protect\citeauthoryear{{Ponman} et~al.}{2003}]{PO03.2}
\begin{barticle}
\bauthor{\binits{T.J.} \bsnm{{Ponman}}},
\bauthor{\binits{A.J.R.} \bsnm{{Sanderson}}},
\bauthor{\binits{A.} \bsnm{{Finoguenov}}},
\batitle{The birmingham-cfa cluster scaling project - iii. entropy and
  similarity in galaxy systems}.
\bjtitle{MNRAS}
\bvolume{343},
\bfpage{331}--\blpage{342}
(\byear{2003}).
doi:\doiurl{10.1046/j.1365-8711.2003.06677.x}
\end{barticle}
\endbibitem

\bibitem[\protect\citeauthoryear{{Postman} et~al.}{2012}]{PO12.1}
\begin{barticle}
\bauthor{\binits{M.} \bsnm{{Postman}}},
\bauthor{\binits{D.} \bsnm{{Coe}}},
\bauthor{\binits{N.} \bsnm{{Ben{\'{\i}}tez}}},
\bauthor{\binits{L.} \bsnm{{Bradley}}},
\bauthor{\binits{T.} \bsnm{{Broadhurst}}},
\bauthor{\binits{M.} \bsnm{{Donahue}}},
\bauthor{\binits{H.} \bsnm{{Ford}}},
\bauthor{\binits{O.} \bsnm{{Graur}}},
\bauthor{\binits{G.} \bsnm{{Graves}}},
\bauthor{\binits{S.} \bsnm{{Jouvel}}},
\bauthor{\binits{A.} \bsnm{{Koekemoer}}},
\bauthor{\binits{D.} \bsnm{{Lemze}}},
\bauthor{\binits{E.} \bsnm{{Medezinski}}},
\bauthor{\binits{A.} \bsnm{{Molino}}},
\bauthor{\binits{L.} \bsnm{{Moustakas}}},
\bauthor{\binits{S.} \bsnm{{Ogaz}}},
\bauthor{\binits{A.} \bsnm{{Riess}}},
\bauthor{\binits{S.} \bsnm{{Rodney}}},
\bauthor{\binits{P.} \bsnm{{Rosati}}},
\bauthor{\binits{K.} \bsnm{{Umetsu}}},
\bauthor{\binits{W.} \bsnm{{Zheng}}},
\bauthor{\binits{A.} \bsnm{{Zitrin}}},
\bauthor{\binits{M.} \bsnm{{Bartelmann}}},
\bauthor{\binits{R.} \bsnm{{Bouwens}}},
\bauthor{\binits{N.} \bsnm{{Czakon}}},
\bauthor{\binits{S.} \bsnm{{Golwala}}},
\bauthor{\binits{O.} \bsnm{{Host}}},
\bauthor{\binits{L.} \bsnm{{Infante}}},
\bauthor{\binits{S.} \bsnm{{Jha}}},
\bauthor{\binits{Y.} \bsnm{{Jimenez-Teja}}},
\bauthor{\binits{D.} \bsnm{{Kelson}}},
\bauthor{\binits{O.} \bsnm{{Lahav}}},
\bauthor{\binits{R.} \bsnm{{Lazkoz}}},
\bauthor{\binits{D.} \bsnm{{Maoz}}},
\bauthor{\binits{C.} \bsnm{{McCully}}},
\bauthor{\binits{P.} \bsnm{{Melchior}}},
\bauthor{\binits{M.} \bsnm{{Meneghetti}}},
\bauthor{\binits{J.} \bsnm{{Merten}}},
\bauthor{\binits{J.} \bsnm{{Moustakas}}},
\bauthor{\binits{M.} \bsnm{{Nonino}}},
\bauthor{\binits{B.} \bsnm{{Patel}}},
\bauthor{\binits{E.} \bsnm{{Reg{\"o}s}}},
\bauthor{\binits{J.} \bsnm{{Sayers}}},
\bauthor{\binits{S.} \bsnm{{Seitz}}},
\bauthor{\binits{A.} \bsnm{{Van der Wel}}},
\batitle{The cluster lensing and supernova survey with hubble: An overview}.
\bjtitle{ApJs}
\bvolume{199},
\bfpage{25}
(\byear{2012}).
doi:\doiurl{10.1088/0067-0049/199/2/25}
\end{barticle}
\endbibitem

\bibitem[\protect\citeauthoryear{{Power} et~al.}{2003}]{PO03.1}
\begin{barticle}
\bauthor{\binits{C.} \bsnm{{Power}}},
\bauthor{\binits{J.F.} \bsnm{{Navarro}}},
\bauthor{\binits{A.} \bsnm{{Jenkins}}},
\bauthor{\binits{C.S.} \bsnm{{Frenk}}},
\bauthor{\binits{S.D.M.} \bsnm{{White}}},
\bauthor{\binits{V.} \bsnm{{Springel}}},
\bauthor{\binits{J.} \bsnm{{Stadel}}},
\bauthor{\binits{T.} \bsnm{{Quinn}}},
\batitle{The inner structure of {$\Lambda$}cdm haloes - i. a numerical
  convergence study}.
\bjtitle{MNRAS}
\bvolume{338},
\bfpage{14}--\blpage{34}
(\byear{2003}).
doi:\doiurl{10.1046/j.1365-8711.2003.05925.x}
\end{barticle}
\endbibitem

\bibitem[\protect\citeauthoryear{{Prada} et~al.}{2012}]{PR12.1}
\begin{barticle}
\bauthor{\binits{F.} \bsnm{{Prada}}},
\bauthor{\binits{A.A.} \bsnm{{Klypin}}},
\bauthor{\binits{A.J.} \bsnm{{Cuesta}}},
\bauthor{\binits{J.E.} \bsnm{{Betancort-Rijo}}},
\bauthor{\binits{J.} \bsnm{{Primack}}},
\batitle{Halo concentrations in the standard {$\Lambda$} cold dark matter
  cosmology}.
\bjtitle{MNRAS}
\bvolume{423},
\bfpage{3018}--\blpage{3030}
(\byear{2012}).
doi:\doiurl{10.1111/j.1365-2966.2012.21007.x}
\end{barticle}
\endbibitem

\bibitem[\protect\citeauthoryear{{Pratt} et~al.}{2007}]{PR07.1}
\begin{barticle}
\bauthor{\binits{G.W.} \bsnm{{Pratt}}},
\bauthor{\binits{H.} \bsnm{{B{\"o}hringer}}},
\bauthor{\binits{J.H.} \bsnm{{Croston}}},
\bauthor{\binits{M.} \bsnm{{Arnaud}}},
\bauthor{\binits{S.} \bsnm{{Borgani}}},
\bauthor{\binits{A.} \bsnm{{Finoguenov}}},
\bauthor{\binits{R.F.} \bsnm{{Temple}}},
\batitle{Temperature profiles of a representative sample of nearby x-ray galaxy
  clusters}.
\bjtitle{A\&A}
\bvolume{461},
\bfpage{71}--\blpage{80}
(\byear{2007}).
doi:\doiurl{10.1051/0004-6361:20065676}
\end{barticle}
\endbibitem

\bibitem[\protect\citeauthoryear{{Pratt} et~al.}{2010}]{PR10.1}
\begin{barticle}
\bauthor{\binits{G.W.} \bsnm{{Pratt}}},
\bauthor{\binits{M.} \bsnm{{Arnaud}}},
\bauthor{\binits{R.} \bsnm{{Piffaretti}}},
\bauthor{\binits{H.} \bsnm{{B{\"o}hringer}}},
\bauthor{\binits{T.J.} \bsnm{{Ponman}}},
\bauthor{\binits{J.H.} \bsnm{{Croston}}},
\bauthor{\binits{G.M.} \bsnm{{Voit}}},
\bauthor{\binits{S.} \bsnm{{Borgani}}},
\bauthor{\binits{R.G.} \bsnm{{Bower}}},
\batitle{Gas entropy in a representative sample of nearby x-ray galaxy clusters
  (rexcess): relationship to gas mass fraction}.
\bjtitle{A\&A}
\bvolume{511},
\bfpage{85}
(\byear{2010}).
doi:\doiurl{10.1051/0004-6361/200913309}
\end{barticle}
\endbibitem

\bibitem[\protect\citeauthoryear{{Rines} and {Diaferio}}{2006}]{RI06.1}
\begin{barticle}
\bauthor{\binits{K.} \bsnm{{Rines}}},
\bauthor{\binits{A.} \bsnm{{Diaferio}}},
\batitle{Cirs: Cluster infall regions in the sloan digital sky survey. i.
  infall patterns and mass profiles}.
\bjtitle{AJ}
\bvolume{132},
\bfpage{1275}--\blpage{1297}
(\byear{2006}).
doi:\doiurl{10.1086/506017}
\end{barticle}
\endbibitem

\bibitem[\protect\citeauthoryear{{S{\'a}nchez-Conde} et~al.}{2011}]{SA11.1}
\begin{barticle}
\bauthor{\binits{M.A.} \bsnm{{S{\'a}nchez-Conde}}},
\bauthor{\binits{M.} \bsnm{{Cannoni}}},
\bauthor{\binits{F.} \bsnm{{Zandanel}}},
\bauthor{\binits{M.E.} \bsnm{{G{\'o}mez}}},
\bauthor{\binits{F.} \bsnm{{Prada}}},
\batitle{Dark matter searches with cherenkov telescopes: nearby dwarf galaxies
  or local galaxy clusters?}
\bjtitle{JCAP}
\bvolume{12},
\bfpage{11}
(\byear{2011}).
doi:\doiurl{10.1088/1475-7516/2011/12/011}
\end{barticle}
\endbibitem

\bibitem[\protect\citeauthoryear{{Sand} et~al.}{2004}]{SA04.1}
\begin{barticle}
\bauthor{\binits{D.J.} \bsnm{{Sand}}},
\bauthor{\binits{T.} \bsnm{{Treu}}},
\bauthor{\binits{G.P.} \bsnm{{Smith}}},
\bauthor{\binits{R.S.} \bsnm{{Ellis}}},
\batitle{The dark matter distribution in the central regions of galaxy
  clusters: Implications for cold dark matter}.
\bjtitle{ApJ}
\bvolume{604},
\bfpage{88}--\blpage{107}
(\byear{2004})
\end{barticle}
\endbibitem

\bibitem[\protect\citeauthoryear{{Schmidt} and {Allen}}{2007}]{SC07.4}
\begin{barticle}
\bauthor{\binits{R.W.} \bsnm{{Schmidt}}},
\bauthor{\binits{S.W.} \bsnm{{Allen}}},
\batitle{The dark matter haloes of massive, relaxed galaxy clusters observed
  with chandra}.
\bjtitle{MNRAS}
\bvolume{379},
\bfpage{209}--\blpage{221}
(\byear{2007}).
doi:\doiurl{10.1111/j.1365-2966.2007.11928.x}
\end{barticle}
\endbibitem

\bibitem[\protect\citeauthoryear{{Schneider} and {Er}}{2008}]{SC08.1}
\begin{barticle}
\bauthor{\binits{P.} \bsnm{{Schneider}}},
\bauthor{\binits{X.} \bsnm{{Er}}},
\batitle{Weak lensing goes bananas: what flexion really measures}.
\bjtitle{A\&A}
\bvolume{485},
\bfpage{363}--\blpage{376}
(\byear{2008}).
doi:\doiurl{10.1051/0004-6361:20078631}
\end{barticle}
\endbibitem

\bibitem[\protect\citeauthoryear{{Seljak}}{2000}]{SE00.1}
\begin{barticle}
\bauthor{\binits{U.} \bsnm{{Seljak}}},
\batitle{Analytic model for galaxy and dark matter clustering}.
\bjtitle{MNRAS}
\bvolume{318},
\bfpage{203}--\blpage{213}
(\byear{2000}).
doi:\doiurl{10.1046/j.1365-8711.2000.03715.x}
\end{barticle}
\endbibitem

\bibitem[\protect\citeauthoryear{{Sereno} and {Zitrin}}{2012}]{SE12.1}
\begin{barticle}
\bauthor{\binits{M.} \bsnm{{Sereno}}},
\bauthor{\binits{A.} \bsnm{{Zitrin}}},
\batitle{Triaxial strong-lensing analysis of the $z > 0.5$ macs clusters: the
  mass-concentration relation}.
\bjtitle{MNRAS}
\bvolume{419},
\bfpage{3280}--\blpage{3291}
(\byear{2012}).
doi:\doiurl{10.1111/j.1365-2966.2011.19968.x}
\end{barticle}
\endbibitem

\bibitem[\protect\citeauthoryear{{Shaw} et~al.}{2006}]{SH06.2}
\begin{barticle}
\bauthor{\binits{L.D.} \bsnm{{Shaw}}},
\bauthor{\binits{J.} \bsnm{{Weller}}},
\bauthor{\binits{J.P.} \bsnm{{Ostriker}}},
\bauthor{\binits{P.} \bsnm{{Bode}}},
\batitle{Statistics of physical properties of dark matter clusters}.
\bjtitle{ApJ}
\bvolume{646},
\bfpage{815}--\blpage{833}
(\byear{2006}).
doi:\doiurl{10.1086/505016}
\end{barticle}
\endbibitem

\bibitem[\protect\citeauthoryear{{Springel} et~al.}{2005}]{SP05.1}
\begin{barticle}
\bauthor{\binits{V.} \bsnm{{Springel}}},
\bauthor{\binits{S.D.M.} \bsnm{{White}}},
\bauthor{\binits{A.} \bsnm{{Jenkins}}},
\bauthor{\binits{C.S.} \bsnm{{Frenk}}},
\bauthor{\binits{N.} \bsnm{{Yoshida}}},
\bauthor{\binits{L.} \bsnm{{Gao}}},
\bauthor{\binits{J.} \bsnm{{Navarro}}},
\bauthor{\binits{R.} \bsnm{{Thacker}}},
\bauthor{\binits{D.} \bsnm{{Croton}}},
\bauthor{\binits{J.} \bsnm{{Helly}}},
\bauthor{\binits{J.A.} \bsnm{{Peacock}}},
\bauthor{\binits{S.} \bsnm{{Cole}}},
\bauthor{\binits{P.} \bsnm{{Thomas}}},
\bauthor{\binits{H.} \bsnm{{Couchman}}},
\bauthor{\binits{A.} \bsnm{{Evrard}}},
\bauthor{\binits{J.} \bsnm{{Colberg}}},
\bauthor{\binits{F.} \bsnm{{Pearce}}},
\batitle{Simulations of the formation, evolution and clustering of galaxies and
  quasars}.
\bjtitle{Nat}
\bvolume{435},
\bfpage{629}--\blpage{636}
(\byear{2005}).
doi:\doiurl{10.1038/nature03597}
\end{barticle}
\endbibitem

\bibitem[\protect\citeauthoryear{{Subramanian} et~al.}{2000}]{SU00.1}
\begin{barticle}
\bauthor{\binits{K.} \bsnm{{Subramanian}}},
\bauthor{\binits{R.} \bsnm{{Cen}}},
\bauthor{\binits{J.P.} \bsnm{{Ostriker}}},
\batitle{The structure of dark matter halos in hierarchical clustering
  theories}.
\bjtitle{ApJ}
\bvolume{538},
\bfpage{528}--\blpage{542}
(\byear{2000}).
doi:\doiurl{10.1086/309152}
\end{barticle}
\endbibitem

\bibitem[\protect\citeauthoryear{{Syer} and {White}}{1998}]{SY98.1}
\begin{barticle}
\bauthor{\binits{D.} \bsnm{{Syer}}},
\bauthor{\binits{S.D.M.} \bsnm{{White}}},
\batitle{Dark halo mergers and the formation of a universal profile}.
\bjtitle{MNRAS}
\bvolume{293},
\bfpage{337}
(\byear{1998}).
doi:\doiurl{10.1046/j.1365-8711.1998.01285.x}
\end{barticle}
\endbibitem

\bibitem[\protect\citeauthoryear{{Umetsu} and {Broadhurst}}{2008}]{UM08.1}
\begin{barticle}
\bauthor{\binits{K.} \bsnm{{Umetsu}}},
\bauthor{\binits{T.} \bsnm{{Broadhurst}}},
\batitle{Combining lens distortion and depletion to map the mass distribution
  of a1689}.
\bjtitle{ApJ}
\bvolume{684},
\bfpage{177}--\blpage{203}
(\byear{2008}).
doi:\doiurl{10.1086/589683}
\end{barticle}
\endbibitem

\bibitem[\protect\citeauthoryear{{Umetsu} et~al.}{2010}]{UM10.1}
\begin{barticle}
\bauthor{\binits{K.} \bsnm{{Umetsu}}},
\bauthor{\binits{E.} \bsnm{{Medezinski}}},
\bauthor{\binits{T.} \bsnm{{Broadhurst}}},
\bauthor{\binits{A.} \bsnm{{Zitrin}}},
\bauthor{\binits{N.} \bsnm{{Okabe}}},
\bauthor{\binits{B.-C.} \bsnm{{Hsieh}}},
\bauthor{\binits{S.M.} \bsnm{{Molnar}}},
\batitle{The mass structure of the galaxy cluster cl0024+1654 from a full
  lensing analysis of joint subaru and acs/nic3 observations}.
\bjtitle{ApJ}
\bvolume{714},
\bfpage{1470}--\blpage{1496}
(\byear{2010}).
doi:\doiurl{10.1088/0004-637X/714/2/1470}
\end{barticle}
\endbibitem

\bibitem[\protect\citeauthoryear{{Umetsu} et~al.}{2011}]{UM11.1}
\begin{barticle}
\bauthor{\binits{K.} \bsnm{{Umetsu}}},
\bauthor{\binits{T.} \bsnm{{Broadhurst}}},
\bauthor{\binits{A.} \bsnm{{Zitrin}}},
\bauthor{\binits{E.} \bsnm{{Medezinski}}},
\bauthor{\binits{D.} \bsnm{{Coe}}},
\bauthor{\binits{M.} \bsnm{{Postman}}},
\batitle{A precise cluster mass profile averaged from the highest-quality
  lensing data}.
\bjtitle{ApJ}
\bvolume{738},
\bfpage{41}
(\byear{2011}).
doi:\doiurl{10.1088/0004-637X/738/1/41}
\end{barticle}
\endbibitem

\bibitem[\protect\citeauthoryear{{Umetsu} et~al.}{2012}]{UM12.1}
\begin{barticle}
\bauthor{\binits{K.} \bsnm{{Umetsu}}},
\bauthor{\binits{E.} \bsnm{{Medezinski}}},
\bauthor{\binits{M.} \bsnm{{Nonino}}},
\bauthor{\binits{J.} \bsnm{{Merten}}},
\bauthor{\binits{A.} \bsnm{{Zitrin}}},
\bauthor{\binits{A.} \bsnm{{Molino}}},
\bauthor{\binits{C.} \bsnm{{Grillo}}},
\bauthor{\binits{M.} \bsnm{{Carrasco}}},
\bauthor{\binits{M.} \bsnm{{Donahue}}},
\bauthor{\binits{A.} \bsnm{{Mahdavi}}},
\bauthor{\binits{D.} \bsnm{{Coe}}},
\bauthor{\binits{M.} \bsnm{{Postman}}},
\bauthor{\binits{A.} \bsnm{{Koekemoer}}},
\bauthor{\binits{N.} \bsnm{{Czakon}}},
\bauthor{\binits{J.} \bsnm{{Sayers}}},
\bauthor{\binits{T.} \bsnm{{Mroczkowski}}},
\bauthor{\binits{S.} \bsnm{{Golwala}}},
\bauthor{\binits{P.M.} \bsnm{{Koch}}},
\bauthor{\binits{K.-Y.} \bsnm{{Lin}}},
\bauthor{\binits{S.M.} \bsnm{{Molnar}}},
\bauthor{\binits{P.} \bsnm{{Rosati}}},
\bauthor{\binits{I.} \bsnm{{Balestra}}},
\bauthor{\binits{A.} \bsnm{{Mercurio}}},
\bauthor{\binits{M.} \bsnm{{Scodeggio}}},
\bauthor{\binits{A.} \bsnm{{Biviano}}},
\bauthor{\binits{T.} \bsnm{{Anguita}}},
\bauthor{\binits{L.} \bsnm{{Infante}}},
\bauthor{\binits{G.} \bsnm{{Seidel}}},
\bauthor{\binits{I.} \bsnm{{Sendra}}},
\bauthor{\binits{S.} \bsnm{{Jouvel}}},
\bauthor{\binits{O.} \bsnm{{Host}}},
\bauthor{\binits{D.} \bsnm{{Lemze}}},
\bauthor{\binits{T.} \bsnm{{Broadhurst}}},
\bauthor{\binits{M.} \bsnm{{Meneghetti}}},
\bauthor{\binits{L.} \bsnm{{Moustakas}}},
\bauthor{\binits{M.} \bsnm{{Bartelmann}}},
\bauthor{\binits{N.} \bsnm{{Ben{\'{\i}}tez}}},
\bauthor{\binits{R.} \bsnm{{Bouwens}}},
\bauthor{\binits{L.} \bsnm{{Bradley}}},
\bauthor{\binits{H.} \bsnm{{Ford}}},
\bauthor{\binits{Y.} \bsnm{{Jim{\'e}nez-Teja}}},
\bauthor{\binits{D.} \bsnm{{Kelson}}},
\bauthor{\binits{O.} \bsnm{{Lahav}}},
\bauthor{\binits{P.} \bsnm{{Melchior}}},
\bauthor{\binits{J.} \bsnm{{Moustakas}}},
\bauthor{\binits{S.} \bsnm{{Ogaz}}},
\bauthor{\binits{S.} \bsnm{{Seitz}}},
\bauthor{\binits{W.} \bsnm{{Zheng}}},
\batitle{Clash: Mass distribution in and around macs j1206.2-0847 from a full
  cluster lensing analysis}.
\bjtitle{ApJ}
\bvolume{755},
\bfpage{56}
(\byear{2012}).
doi:\doiurl{10.1088/0004-637X/755/1/56}
\end{barticle}
\endbibitem

\bibitem[\protect\citeauthoryear{{Vazza} et~al.}{2012}]{VA12.1}
\begin{botherref}
\oauthor{\binits{F.} \bsnm{{Vazza}}},
\oauthor{\binits{M.} \bsnm{{Br{\"u}ggen}}},
\oauthor{\binits{C.} \bsnm{{Gheller}}},
Thermal and non-thermal traces of agn feedback: results from cosmological amr
  simulations.
MNRAS,
194
(2012).
doi:\doiurl{10.1093/mnras/sts213}
\end{botherref}
\endbibitem

\bibitem[\protect\citeauthoryear{{Velander} et~al.}{2011}]{VE11.1}
\begin{barticle}
\bauthor{\binits{M.} \bsnm{{Velander}}},
\bauthor{\binits{K.} \bsnm{{Kuijken}}},
\bauthor{\binits{T.} \bsnm{{Schrabback}}},
\batitle{Probing galaxy dark matter haloes in cosmos with weak lensing
  flexion}.
\bjtitle{MNRAS}
\bvolume{412},
\bfpage{2665}--\blpage{2677}
(\byear{2011}).
doi:\doiurl{10.1111/j.1365-2966.2010.18085.x}
\end{barticle}
\endbibitem

\bibitem[\protect\citeauthoryear{{Vikhlinin} et~al.}{2005}]{VI05.1}
\begin{barticle}
\bauthor{\binits{A.} \bsnm{{Vikhlinin}}},
\bauthor{\binits{M.} \bsnm{{Markevitch}}},
\bauthor{\binits{S.S.} \bsnm{{Murray}}},
\bauthor{\binits{C.} \bsnm{{Jones}}},
\bauthor{\binits{W.} \bsnm{{Forman}}},
\bauthor{\binits{L.} \bsnm{{Van Speybroeck}}},
\batitle{Chandra temperature profiles for a sample of nearby relaxed galaxy
  clusters}.
\bjtitle{ApJ}
\bvolume{628},
\bfpage{655}--\blpage{672}
(\byear{2005}).
doi:\doiurl{10.1086/431142}
\end{barticle}
\endbibitem

\bibitem[\protect\citeauthoryear{{Vikhlinin} et~al.}{2006}]{VI06.1}
\begin{barticle}
\bauthor{\binits{A.} \bsnm{{Vikhlinin}}},
\bauthor{\binits{A.} \bsnm{{Kravtsov}}},
\bauthor{\binits{W.} \bsnm{{Forman}}},
\bauthor{\binits{C.} \bsnm{{Jones}}},
\bauthor{\binits{M.} \bsnm{{Markevitch}}},
\bauthor{\binits{S.S.} \bsnm{{Murray}}},
\bauthor{\binits{L.} \bsnm{{Van Speybroeck}}},
\batitle{Chandra sample of nearby relaxed galaxy clusters: Mass, gas fraction,
  and mass-temperature relation}.
\bjtitle{ApJ}
\bvolume{640},
\bfpage{691}--\blpage{709}
(\byear{2006}).
doi:\doiurl{10.1086/500288}
\end{barticle}
\endbibitem

\bibitem[\protect\citeauthoryear{{Viola} et~al.}{2012}]{VI12.1}
\begin{barticle}
\bauthor{\binits{M.} \bsnm{{Viola}}},
\bauthor{\binits{P.} \bsnm{{Melchior}}},
\bauthor{\binits{M.} \bsnm{{Bartelmann}}},
\batitle{Shear-flexion cross-talk in weak-lensing measurements}.
\bjtitle{MNRAS}
\bvolume{419},
\bfpage{2215}--\blpage{2225}
(\byear{2012}).
doi:\doiurl{10.1111/j.1365-2966.2011.19872.x}
\end{barticle}
\endbibitem

\bibitem[\protect\citeauthoryear{{Voigt} and {Fabian}}{2006}]{VO06.1}
\begin{barticle}
\bauthor{\binits{L.M.} \bsnm{{Voigt}}},
\bauthor{\binits{A.C.} \bsnm{{Fabian}}},
\batitle{Galaxy cluster mass profiles}.
\bjtitle{MNRAS}
\bvolume{368},
\bfpage{518}--\blpage{533}
(\byear{2006}).
doi:\doiurl{10.1111/j.1365-2966.2006.10199.x}
\end{barticle}
\endbibitem

\bibitem[\protect\citeauthoryear{{Voigt} et~al.}{2002}]{VO02.1}
\begin{barticle}
\bauthor{\binits{L.M.} \bsnm{{Voigt}}},
\bauthor{\binits{R.W.} \bsnm{{Schmidt}}},
\bauthor{\binits{A.C.} \bsnm{{Fabian}}},
\bauthor{\binits{S.W.} \bsnm{{Allen}}},
\bauthor{\binits{R.M.} \bsnm{{Johnstone}}},
\batitle{Conduction and cooling flows}.
\bjtitle{MNRAS}
\bvolume{335},
\bfpage{7}--\blpage{11}
(\byear{2002}).
doi:\doiurl{10.1046/j.1365-8711.2002.05741.x}
\end{barticle}
\endbibitem

\bibitem[\protect\citeauthoryear{{Voit} et~al.}{2003}]{VO03.1}
\begin{barticle}
\bauthor{\binits{G.M.} \bsnm{{Voit}}},
\bauthor{\binits{M.L.} \bsnm{{Balogh}}},
\bauthor{\binits{R.G.} \bsnm{{Bower}}},
\bauthor{\binits{C.G.} \bsnm{{Lacey}}},
\bauthor{\binits{G.L.} \bsnm{{Bryan}}},
\batitle{On the origin of intracluster entropy}.
\bjtitle{ApJ}
\bvolume{593},
\bfpage{272}--\blpage{290}
(\byear{2003}).
doi:\doiurl{10.1086/376499}
\end{barticle}
\endbibitem

\bibitem[\protect\citeauthoryear{{Wang} and {White}}{2009}]{WA09.1}
\begin{barticle}
\bauthor{\binits{J.} \bsnm{{Wang}}},
\bauthor{\binits{S.D.M.} \bsnm{{White}}},
\batitle{Are mergers responsible for universal halo properties?}
\bjtitle{MNRAS}
\bvolume{396},
\bfpage{709}--\blpage{717}
(\byear{2009}).
doi:\doiurl{10.1111/j.1365-2966.2009.14755.x}
\end{barticle}
\endbibitem

\bibitem[\protect\citeauthoryear{{Wojtak} and {{\L}okas}}{2010}]{WO10.1}
\begin{barticle}
\bauthor{\binits{R.} \bsnm{{Wojtak}}},
\bauthor{\binits{E.L.} \bsnm{{{\L}okas}}},
\batitle{Mass profiles and galaxy orbits in nearby galaxy clusters from the
  analysis of the projected phase space}.
\bjtitle{MNRAS}
\bvolume{408},
\bfpage{2442}--\blpage{2456}
(\byear{2010}).
doi:\doiurl{10.1111/j.1365-2966.2010.17297.x}
\end{barticle}
\endbibitem

\bibitem[\protect\citeauthoryear{{Yuan} et~al.}{2010}]{YU10.1}
\begin{barticle}
\bauthor{\binits{Q.} \bsnm{{Yuan}}},
\bauthor{\binits{P.-F.} \bsnm{{Yin}}},
\bauthor{\binits{X.-J.} \bsnm{{Bi}}},
\bauthor{\binits{X.-M.} \bsnm{{Zhang}}},
\bauthor{\binits{S.-H.} \bsnm{{Zhu}}},
\batitle{Gamma rays and neutrinos from dark matter annihilation in galaxy
  clusters}.
\bjtitle{PRD}
\bvolume{82}(\bissue{2}),
\bfpage{023506}
(\byear{2010}).
doi:\doiurl{10.1103/PhysRevD.82.023506}
\end{barticle}
\endbibitem

\bibitem[\protect\citeauthoryear{{Zentner} et~al.}{2005}]{ZE05.1}
\begin{barticle}
\bauthor{\binits{A.R.} \bsnm{{Zentner}}},
\bauthor{\binits{A.A.} \bsnm{{Berlind}}},
\bauthor{\binits{J.S.} \bsnm{{Bullock}}},
\bauthor{\binits{A.V.} \bsnm{{Kravtsov}}},
\bauthor{\binits{R.H.} \bsnm{{Wechsler}}},
\batitle{The physics of galaxy clustering. i. a model for subhalo populations}.
\bjtitle{ApJ}
\bvolume{624},
\bfpage{505}--\blpage{525}
(\byear{2005}).
doi:\doiurl{10.1086/428898}
\end{barticle}
\endbibitem

\bibitem[\protect\citeauthoryear{{Zhao} et~al.}{2009}]{ZH09.1}
\begin{barticle}
\bauthor{\binits{D.H.} \bsnm{{Zhao}}},
\bauthor{\binits{Y.P.} \bsnm{{Jing}}},
\bauthor{\binits{H.J.} \bsnm{{Mo}}},
\bauthor{\binits{G.} \bsnm{{B{\"o}rner}}},
\batitle{Accurate universal models for the mass accretion histories and
  concentrations of dark matter halos}.
\bjtitle{ApJ}
\bvolume{707},
\bfpage{354}--\blpage{369}
(\byear{2009}).
doi:\doiurl{10.1088/0004-637X/707/1/354}
\end{barticle}
\endbibitem

\end{thebibliography}

\end{document}